\documentclass{aa}  

\usepackage{graphicx}
\usepackage{txfonts}
\newcommand{\doce}{\mbox{$^{12}$CO}}
\newcommand{\dsiete}{\mbox{C$^{17}$O}}

\newcommand{\trece}{\mbox{$^{13}$CO}}

\newcommand{\htrececn}{\mbox{H$^{13}$CN}}
\newcommand{\jsc}{\mbox{$J$=6$-$5}}
\newcommand{\jct}{\mbox{$J$=4$-$3}}
\newcommand{\jtd}{\mbox{$J$=3$-$2}}
\newcommand{\jdu}{\mbox{$J$=2$-$1}}
\newcommand{\juc}{\mbox{$J$=1$-$0}}
\newcommand{\jdn}{\mbox{$J$=10$-$9}}
\newcommand{\jdq}{\mbox{$J$=16$-$15}}

\newcommand{\kms}{\mbox{km\,s$^{-1}$}}

\newcommand{\ms}{\mbox{$M_{\mbox{\sun}}$}}

\newcommand{\my}{\mbox{$M_{\mbox{\sun}}$\,yr$^{-1}$}}

\newcommand{\mloss}{\mbox{$\dot{M}$}}

\newcommand{\lsim}{\raisebox{-.4ex}{$\stackrel{\sf <}{\scriptstyle\sf \sim}$}}
\newcommand{\gsim}{\raisebox{-.4ex}{$\stackrel{\sf >}{\scriptstyle\sf \sim}$}}

\newcommand{\farcss}{\mbox{\rlap{.}$''$}}
\newcommand{\fdeg}{\mbox{\rlap{.}$^\circ$}}
\begin{document}

   \title{Further ALMA observations and detailed modeling of the Red Rectangle}

   \author{V. Bujarrabal
          \inst{1}
          \and
          A. Castro-Carrizo\inst{2}
            \and J. Alcolea\inst{3}  
          \and
M. Santander-Garc\'{i}a\inst{3,4} \and
H. Van Winckel\inst{5}  \and C. S\'anchez Contreras\inst{6}  
          }

   \institute{             Observatorio Astron\'omico Nacional (OAN-IGN),
              Apartado 112, E-28803 Alcal\'a de Henares, Spain\\
              \email{v.bujarrabal@oan.es}
\and 
 Institut de Radioastronomie Millim\'etrique, 300 rue de la Piscine,
 38406, Saint Martin d'H\`eres, France  
\and
             Observatorio Astron\'omico Nacional (OAN-IGN),
             C/ Alfonso XII, 3, E-28014 Madrid, Spain
             \and
             Instituto de Ciencia de Materiales de Madrid (CSIC). Calle
             Sor Juana In\'es de la Cruz 3, E-28049 Cantoblanco, Madrid,
             Spain 
\and
Instituut voor Sterrenkunde, K.U.Leuven, Celestijnenlaan 200B, 3001
Leuven, Belgium
\and 
Centro de Astrobiolog\'{\i}a (CSIC-INTA), ESAC Campus, E-28691
Villanueva de la Ca\~nada, Madrid, Spain 
           }

   \date{submitted 18/03/2016; accepted 07/06/2016}

  \abstract
   {}
{We aim to study the rotating and expanding gas in the Red Rectangle,
  which is a
  well known object that recently left the asymptotic
   giant branch (AGB) phase. We analyze
   the properties of both components and the relation between
   them.
   Rotating disks have been very elusive in post-AGB nebulae, in
   which gas is almost always found to be in expansion.
}
{We present new high-quality ALMA observations of \dsiete\ \jsc\ and
  \htrececn\ \jct\ line emission and results from a new reduction of
  already published \trece\ \jtd\ data. A detailed model fitting of all
  the molecular line data, including previous maps and single-dish
  observations of lines of CO, CII, and CI, was performed using a
  sophisticated code that includes an accurate nonlocal treatment of
  radiative transfer in 2D. These observations (of low- and
  high-opacity lines requiring various degrees of excitation) and the
  corresponding modeling allowed us to deepen the analysis of the
  nebular properties. We also stress the uncertainties, particularly in
  the determination of the boundaries of the CO-rich gas and some
  properties of the outflow.}
{We confirm the presence of a rotating equatorial disk and an outflow,
which is mainly formed of gas leaving the disk.  The mass of the disk is
  $\sim$ 0.01 \ms, and that of the CO-rich outflow is around ten times
  smaller. High temperatures of \gsim\ 100 K are derived for most
  components. From comparison of the mass values, we roughly estimate
  the lifetime of the rotating disk, which is found to be of
about 10000 yr. Taking data of a few other post-AGB composite nebulae
into account, we find that the lifetimes of disks around post-AGB stars
typically range between 5000 and more than 20000 yr. The angular
momentum of the disk is found to be high, $\sim$ 9 \ms\,AU\,\kms, which
is comparable to that of the stellar system at present. Our
observations of \htrececn\ show a particularly wide velocity dispersion
and indicate that this molecule is only abundant in the inner Keplerian
disk, at \lsim\ 60 AU from the stellar system. We suggest that HCN is
formed in a dense photodissociation region (PDR) due to the UV excess
known to be produced by the stellar system, following chemical
mechanisms that are well established for interstellar medium PDRs and
disks orbiting young stars. We further suggest that this UV excess
could lead to an efficient formation and excitation of PAHs and other
C-bearing macromolecules, whose emission is very intense in the optical
counterpart.  }
  {}


   \maketitle
%

\section{Introduction}

Most planetary and preplanetary nebulae (PNe, PPNe) show conspicuous
departures from spherical symmetry, often with a clear axis of
symmetry. In addition, very massive ($\sim$ 0.1 \ms) and fast (20--200
\kms) bipolar outflows are characteristic of PPNe and thought to be
crucial to understand the formation and shaping of PNe; see, e.g.,
Bujarrabal et al.\ (2001), Balick \& Frank (2002), and S\'anchez
Contreras \& Sahai (2012).  On the other hand, circumstellar envelopes
around asymptotic giant branch (AGB) stars, from which PNe have
evolved, are in general spherical and expand isotropically at moderate
velocities $\sim$ 10--15 \kms; it is thought that their ejection is
powered by radiation pressure. The linear momentum carried by
the post-AGB outflows is too high to be powered by momentum transfer
from stellar photons, which strengthens the need of some other
mechanism to explain the dynamics in this phase.  It is often assumed
that this spectacular evolution is driven by magnetocentrifugal
launching, as probably occurs in forming stars, which implies that
rotating disks must be systematically formed from previously ejected
material (e.g., Soker 2001; Frank \& Blackman 2004). The presence of a
stellar or substellar companion is necessary in nebulae around evolved
stars, since otherwise the ejected shells do not have enough angular
momentum to form such disks.

However, in most planetary and preplanetary nebulae only gas in
expansion is detected (Bujarrabal et al.\ 2001; S\'anchez Contreras
\& Sahai 2012, etc) and the disks postulated to orbit post-AGB stars
have been very elusive. To date, the Keplerian velocity field has been
unambiguously identified by means of interferometric mm wave maps
of CO line emission in only two of these putative disks: in the Red
Rectangle (Bujarrabal et al.\ 2005, 2013b) and AC Her (Bujarrabal et
al.\ 2015).

The Red Rectangle and AC Her belong to a class of binary post-AGB stars
with low-mass nebulae and indications of compact disks (e.g., Van
Winckel 2003); about 100 post-AGB objects in our Galaxy are classified
into this wide class of objects.  These objects are characterized by
their spectral energy distributions (SEDs), which particularly shows a
near-infrared (NIR) excess indicating hot dust close to the stellar
system.  This indeed suggests a stable structure, as the stars are no
longer in a state of copious mass loss (De Ruyter et al.\ 2006; Gezer
et al.\ 2015). The inner dust around a few of these post-AGB stars has
been resolved with optical interferometry confirming the very compact
nature of the emitting region (Deroo et al.\ 2006; Hillen et al.\ 2015,
2016).  Additional indications of the longevity of these disks come
from studies of the high degree of dust processing (e.g., Gielen et
al.\ 2011).  Single-dish observations of \doce\ and \trece\ mm wave
emission in a sample of these post-AGB stars systematically yielded
characteristic line profiles, with a prominent single or double peak
and moderate-velocity wings, which are strikingly similar to those of
the Red Rectangle and AC Her (Bujarrabal et al.\ 2013a). As discussed
in that work, profiles of this kind are also found in disks around
young stars (particularly in T Tauri variables), and have been proven,
both from theoretical and observational grounds, to be very reliable
indicators of rotating disks. It was proposed that Keplerian disks are
relatively widespread in post-AGB nebulae, at least in those
surrounding this class of binary stars.


The Red Rectangle is the prototype and best-studied example of this
wide class of post-AGB nebulae. It consists of an equatorial 
disk orbiting a double stellar system, with a period of about 320 days,
plus a beautiful X-shaped axisymmetric nebula in expansion that is
seen in visible wavelengths (e.g., Men'shchikov et al.\ 2002; Cohen et
al.\ 2004).  The nebula equator and axis of symmetry  are
  easily identifiable in the optical images; the axis is slightly out of
  the plane of the sky (by about 5$^\circ$) and its projection in the
  plane of the sky shows a PA $\sim$ 10--15 degrees. Men'shchikov et
al.\ deduced a distance of 710 pc for this source, which is compatible
with our previous analysis of CO data.

The first maps of the CO mm wave emission from the rotating disk in the
Red Rectangle were obtained with the Plateau de Bure Interferometer
(PdBI; Bujarrabal et al.\ 2005).  The Red Rectangle was also observed
in sub-mm line and continuum emissions using ALMA. Maps of
\doce\ \jtd\ and \jsc\ and \trece\ \jtd\ were presented by Bujarrabal
et al.\ (2013b), together with a simplified modeling of the data. The
rotation of the equatorial disk was conspicuous in those maps. The
position-velocity diagrams along the direction of the equator, PA
$\sim$ 102\fdeg 5, clearly show the signature of Keplerian rotation. In
the ALMA maps by Bujarrabal et al.\ (2013b), the disk appears as an
equatorial feature at low LSR velocities within $\pm$2.5 \kms.
Molecular gas in slow expansion was tentatively detected in the PdBI
and single-dish data and spectacularly confirmed by the ALMA
observations, corresponding to the high-latitude CO emission in the
maps at moderate velocities, which is more or less coincident with the
optical image. This composite structure was confirmed by comparison
with nebula models containing both gas in rotation and expansion
(Bujarrabal et al.\ 2013b). The diameter of the detected disk is $\sim$
5 10$^{16}$ cm ($\sim$ 5$''$) and the extent of the CO-rich outflow is
slightly larger.  In view of the structure and velocity of the
expanding component, which seems in some way an extension of the
rotating disk and roughly occupies the region between the disk and the
X-shaped optical image, the outflowing gas seems to be extracted from
the disk. Such an ejection could be due to interaction with the axial
jet that is active in this object (Witt et al.\ 2009; Thomas et
al.\ 2013) or to low-velocity magnetocentrifugal launching from the
extended disk, which is similar to those present in disks around young
stars (Ferreira et al.\ 2006; Panoglou et al.\ 2012, etc).  In
addition, single-dish data of the Red Rectangle in various transitions
of \doce\ and \trece\ (up to \jdq) also exist and have been analyzed
using sophisticated nonlocal treatments of radiative transfer and line
excitation (Bujarrabal \& Alcolea 2013).

As mentioned, another similar source, AC Her, was later confirmed to
show a Keplerian disk; remarkably, no sign of expansion was found in
this source. Molecular gas in expansion has been found from direct
mapping of CO emission in two of these objects: 89 Her (Bujarrabal et
al.\ 2007) and IRAS\,19125+0343 (unpublished PdBI observations).
Rotation was probably present in the detected central condensation, but
the velocity field was not angularly resolved and those disks must be
significantly smaller than for the Red Rectangle.

The CO emission maps show that the optical X-shaped structure in the
Red Rectangle very probably represents the inner surface of the
biconical (outflowing) component that emits in CO lines, although the
optical image is significantly more extended than the detected CO-rich
gas. This optical X-like feature is composed of emission of PAHs
(polycyclic aromatic hydrocarbons) and ERE features (extended red
emission; also probably due to carbonaceous macromolecules), while
scattered light or atomic emission contribute negligibly to this
feature; see Cohen et al.\ (2004). C-bearing macromolecules must be
particularly abundant in this conical structure, although the disk dust
is silicate-rich.  This property was invoked to conclude that the
central star recently changed from O- to C-rich
owing to nuclear processing in the stellar interior (Waters et
al.\ 1998) or to changes in abundances due to reaccretion of
circumstellar material after dust formation (Van Winckel 2014).  As we
show in what follows, the different composition of both components
could also be due to the development of a PDR in the inner disk.

In this paper, we present unpublished ALMA data of C$^{17}$O \jsc\ and
H$^{13}$CN \jct\ emission, together with a new analysis of some of the
previously published ALMA maps and a detailed modeling of the whole set
of observations of (light) molecules using the sophisticated code
presented in a previous paper (Bujarrabal \& Alcolea 2013). Some
Herschel observations of CII and CI far-infrared (FIR) lines are also
discussed in comparison with the emission of C$^{17}$O and
H$^{13}$CN. 

\section{Observations}

\subsection{New ALMA observations and new reduction of the
  \trece\ \jtd\ maps}

We observed the Red Rectangle with ALMA\footnote{ALMA is a partnership
  of ESO (representing its member states), NSF (USA) and NINS (Japan),
  together with NRC (Canada) and NSC and ASIAA (Taiwan), in cooperation
  with the Republic of Chile. The Joint ALMA Observatory is operated by
  ESO, AUI/NRAO and NAOJ. We made use of the ALMA dataset
  ADS/JAO.ALMA\#2011.0.00223.S. } using receiver bands 7 and 9 in
October-November 2012. Details on the observations and data reduction
can be found in Bujarrabal et al.\ (2013b). Maps of \doce\ and
\trece\ \jtd\ emission (at 0.8 mm, ALMA band 7) and of \doce\ \jsc\ (at
0.4 mm, ALMA band 9) were already published (Bujarrabal et
al.\ 2013b). Here we present observations of the \htrececn\ \jct\ line
(band 7) and the \dsiete\ \jsc\ line (band 9), obtained during the same
observational runs, as well as a new image synthesis for the
\trece\ \jtd\ emission with higher signal-to-noise (S/N) ratio, which
allows a deeper analysis of the outflowing gas.

\begin{figure*}
   \centering
   \includegraphics[width=17.3cm]{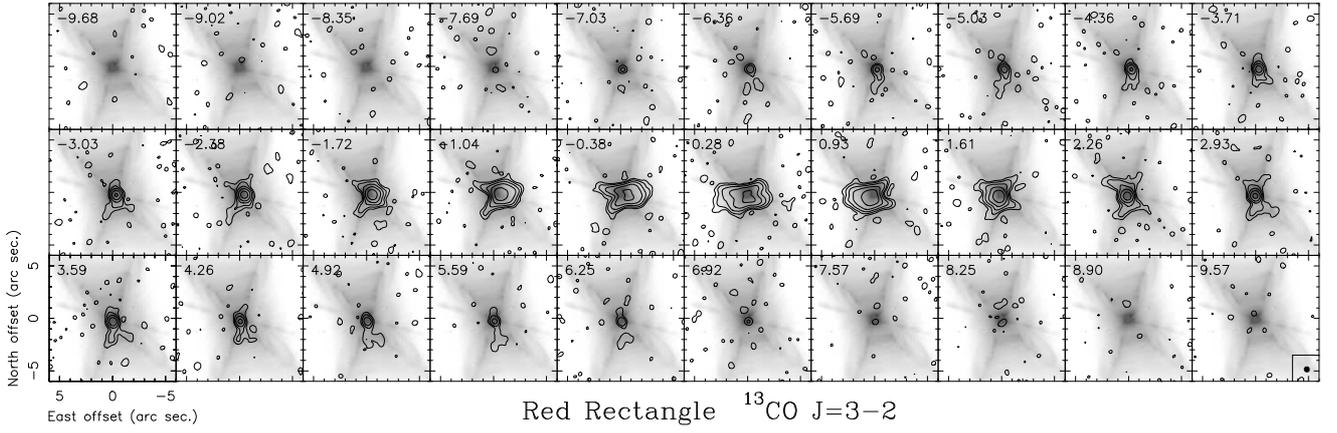}
      \caption{ALMA \trece\ \jtd\ observations of the Red Rectangle. To
        better show the contribution of the various components, these
        maps were obtained using natural weighting in the image
        deconvolution and a velocity resolution of 0.66 \kms. The
        resulting beam size (HPBW) is 0\farcss 59$\times$0\farcss
        56 (shown in the insert). The contours are -0.01, 0.01, 0.03,
        0.09, 0.27, and 0.81 
        Jy/beam (logarithmic spacing by a factor 3, equivalent to 0.34,
        1.02, 3.06, 9.18, and 27.5 K). We note the large contrast between
        expanding gas emission and the peak of the equatorial disk
        brightness, almost by a factor 100. The HST optical image is
        also shown. As in most maps shown in this paper, we
        subtracted the continuum.}
         \label{1}
   \end{figure*}

\begin{figure*}
   \centering
   \includegraphics[width=17.3cm]{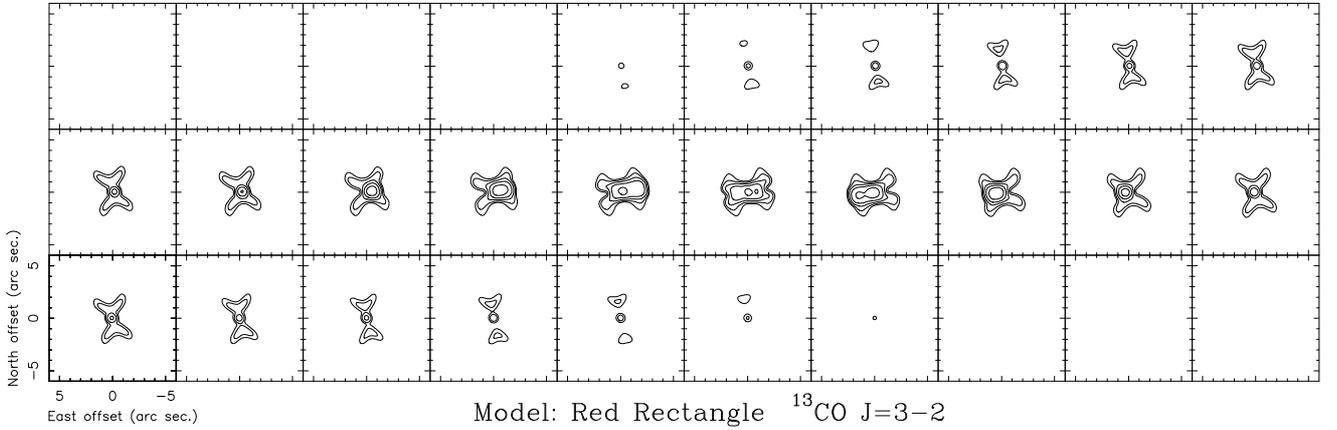}
      \caption{Theoretical maps of \trece\ \jtd\ obtained for our
        best-fit model, to be compared with ALMA maps,
        Fig.\ \ref{1}. The contours and angular and velocity units are
        the same as for the observational data.  }
         \label{2}
   \end{figure*}

In band 9 data and the maps of the weak \htrececn\ line, we used
natural weighting to improve the mapping quality. In our new reduction
of the \trece\ \jtd\ emission, we also used natural weighting. The
half-power beamwidth (HPBW) in band 7 data is of about 0\farcss
59$\times$0\farcss 56 for natural weighting. A HPBW resolution of
$\sim$ 0\farcss 31$\times$0\farcss 25 was obtained in band 9. Our new
results are shown in Figs.\ \ref{1}, \ref{3}, and \ref{5}
(Figs.\ \ref{2}, \ref{4}, and \ref{6} show, respectively, our best
model fits of those maps; see Sect.\ 5). In these figures we also show
a Hubble Space Telescope (HST) optical image (see Bujarrabal et
al.\ 2013b for details on the comparison of both images). In all cases
the (logarithmic) contour spacing is given both in units of Jy/beam and
K (Rayleigh-Jeans equivalent brightness temperature); we 
corrected a minor error in the conversion between K and Jy/beam given
in our previous work for the \trece\ \jtd\ maps due to confusion
between natural and robust beam widths.  As shown in Bujarrabal et
al.\ (2013b), continuum emission was detected and resolved, with a
total flux of $\sim$ 0.6 Jy in band 7 and of $\sim$ 3.5 Jy in band
9. To best analyze the weak line emission, the continuum emission was
subtracted from all the channel maps here presented.

\subsection{Herschel/HIFI observations of atomic lines in the Red Rectangle}

The HIFI heterodyne spectrometer on board the Herschel Space Telescope
(Pilbratt et al.\ 2010; de Graauw et al.\ 2010) was used to observe
three carbon transitions. The data were taken in double side-band (DSB)
fast-chop mode, and using the H and V receivers, which were averaged
once ruling out significant differences in the observed lines with both
receivers.

The CI $J$=1$-$0 line (492.161 GHz) was observed on September 07,
2012, in the 1a band, with a spectral resolution of 0.61
\kms. The CI $J$=2$-$1 line (809.344 GHz) was observed on March 23,
2013, in the 3a band and with a spectral resolution of 0.37 \kms. Finally, 
CII $J$=2$-$1 (1901.280 GHz) was secured on March 20, 2013, in the 7b
band, with a spectral resolution of 0.32 \kms. The beam sizes were
43.1, 26.2, and 11.2 arcsec, respectively.

We reduced the data with the standard HIFI pipeline in the HIPE
software with a modified version of the level 2 algorithm that yields
unaveraged spectra with all spectrometer sub-bands stitched
together. The spectra were then exported to CLASS using the hiClass
tool within HIPE, for further inspection. Data suffering from
significant ripple residuals were filtered and discarded and, in each
case,  the
remaining spectra were averaged. Baseline removal of degree 2
was applied to the final spectra. The data were originally calibrated
in antenna temperature units and later converted into main-beam
temperatures (T$_\mathrm{mb}$) and flux units; see more details in, for
example, Alcolea et al.\ (2013). 

\subsection{Flux lost by interferometric over-resolution}

Given the very high resolution of our ALMA data and that some of the
components we detect, notably the outflow, are relatively extended, we
can expect that a fraction of the flux can be lost in the
interferometric process (particularly in these Cycle-0
observations). It is sometimes difficult to estimate the amount of lost
flux, which depends on the coverage of the interferometric visibilities
in the $uv$ plane, on the calibration accuracy, and on the intrinsic
extent and intensity of the brightness distribution.  However, the
amount of lost flux is very relevant in this paper because model
predictions are compared with maps and single-dish observations of
various lines. Single-dish data are available for the
\doce\ \jsc\ line, observed with Herschel/HIFI; see Bujarrabal \&
Alcolea (2013). The angle-integrated flux profile obtained from our
\doce\ \jsc\ map reaches a peak of 70 Jy, therefore the flux recovered
by the ALMA maps is typically about 60--70\% of the total flux (see
App.\ A, Fig.\ A.1). This factor is reasonable, mainly for mm or sub-mm
interferometric observations, and strongly suggests that the
representation of the nebula shape by means of our ALMA maps is
reliable. The percentage of lost flux is higher in the line wings, for
LSR velocities $\sim$ $\pm$5--8 \kms, in which as much as $\sim$ 60\%
of the total emission can be lost.  This is likely resulting from  the
relatively wide and weak emission from the outflow, and we expect it to
be noticeable even at lower frequencies. The lost flux is similar in
the very center of the line, $V_{\rm lsr}$ $\sim$ 0 \kms, probably
because of the extended emission coming from the outflowing gas placed
in the plane of the sky and from the outermost disk regions.

Since the distributions of the \dsiete\ and \htrececn\ emissions are
significantly more compact than that of \doce\ \jsc, a very moderate
amount of flux is expected to be filtered out in these
cases. \trece\ \jtd\ emission is also relatively compact and should
show low over-resolution effects, except, perhaps, for the weak outflow
emission at intermediate velocities. We also expect a moderate flux loss
for the intense \doce\ \jtd\ line (maps presented in Bujarrabal et
al.\ 2013b) because its extent/resolution ratio is smaller than for
\doce\ \jsc\ and the calibration conditions are better for its lower
frequency.  The comparison with the single-dish spectrum published by
De Beck et al. (2010) is not straightforward because of the uncertain
conversion to flux in Jy, but our conclusions seem confirmed with a 
low flux loss in the central part of the profile and a moderate loss of
about 25--30\% in the line wings.

In our data fitting, we try to take into account that the fraction of
lost flux is not always negligible. As seen below, we attempt reach a
compromise between the fittings of the maps and the total intensity
obtained from single-dish data.

   \begin{figure*}
   \centering
   \includegraphics[width=17.3cm]{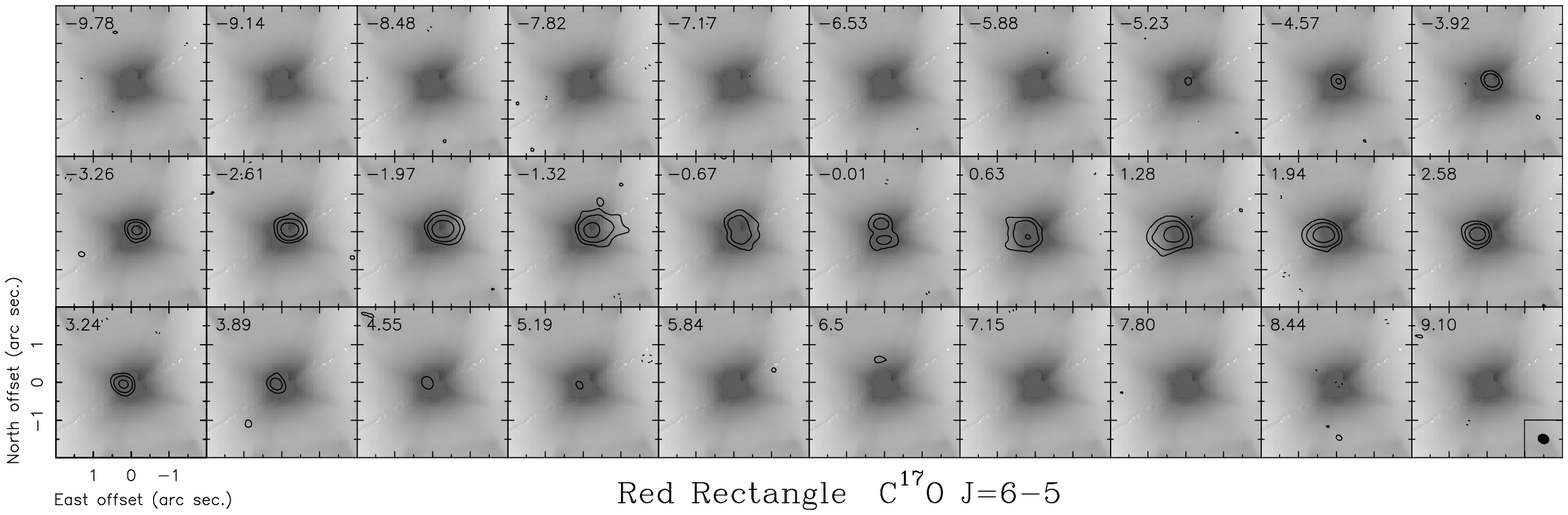}
      \caption{ALMA \dsiete\ \jsc\ observations of the Red
        Rectangle. The resulting beam size (HPBW) is 0\farcss
        31$\times$0\farcss 26 (shown in the insert) and the chosen
        velocity resolution is 0.45 \kms. The contours are -0.075,
        0.075, 0.225, and 0.675 Jy/beam (logarithmic spacing by a
        factor 3, equivalent to 2.54, 7.62, and 22.9 K). The
        emission is weak compared with the
        \doce\ \jsc\ emission shown in Bujarrabal et al.\ (2013b) and
        Fig.\ B.1, which
        reaches a peak of $\sim$ 200 K.  The continuum has been subtracted.}
         \label{3}
   \end{figure*}

   \begin{figure*}
   \centering
   \includegraphics[width=17.3cm]{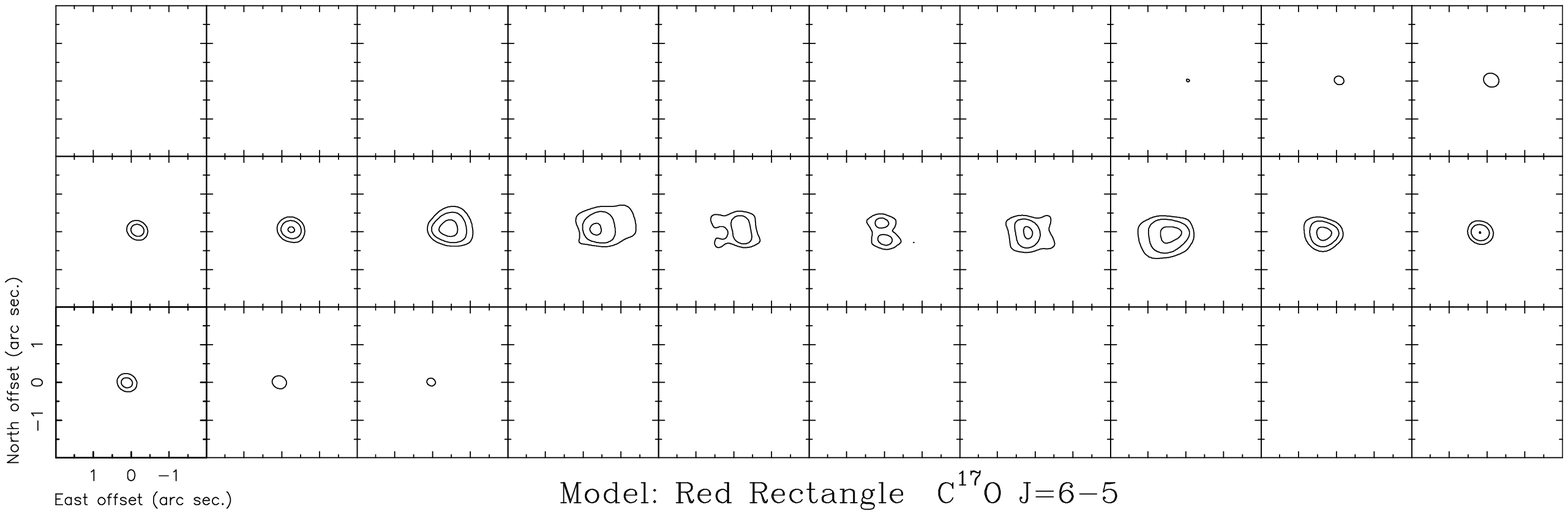}
      \caption{Theoretical maps of \dsiete\ \jsc\ obtained for our
        best-fit model for comparison with the ALMA maps in 
        Fig.\ \ref{3}. The contours and angular and velocity units are
        the same as for the observational data.}
         \label{4}
   \end{figure*}

\section{Line emission model}

We modeled our ALMA maps using a sophisticated and accurate code, which 
simultaneously calculates the molecular level populations and solves the
radiative transfer equations for the geometry and kinematics of the Red
Rectangle. The code was first presented in Bujarrabal \& Alcolea
(2013), which includes the details of the numerical
treatment and exhaustive tests of the calculation accuracy (in
particular for the case of CO). Our model assumes a 2D gas cloud
showing axial symmetry. Radiative transfer is solved with a fully
nonlocal treatment and the molecular level populations are calculated
in a high number of cells. The level populations in each point depend
on the intensity of radiation arriving at the frequencies of the
various transitions and, at the same time, the radiative transfer at
these frequencies depends on the populations of the levels joined by
the transitions. Therefore, both the calculations of the level
populations and radiation intensity are performed simultaneous and
coherently, in an iterative process that involves radiative coupling
between the different parts of the cloud. Collisional transitions are
also taken into account in each cell, and collisional rates were taken from
the LAMBDA database (http://www.strw.leidenuniv.nl/$\sim$moldata).

Finally, once convergence has been attained, the radiative transfer
equations are solved in the direction of the observer. In Bujarrabal \&
Alcolea (2013), the resulting brightness distribution in the plane of
the sky is convolved with single-dish, wide beams, yielding spectral
profiles of the whole emission of the source. Here, the brightness
distribution is convolved with the interferometric beam to obtain
synthetic maps directly comparable to the observed maps. The beam
convolution and calculation of synthetic maps are similar to those
shown in Bujarrabal et al.\ (2013b).

In the preliminary analysis performed by Bujarrabal et al.\ (2013b),
the molecular excitation treatment was much simpler, just assuming LTE
excitation. That approach may be valid for low-$J$ transitions of
\doce, for which the low Einstein coefficients and high optical depth
guarantee thermalization under general conditions. In the
calculations performed in this paper, including CO high-$J$ levels,
optically thin lines, and \htrececn\ transitions, we cannot assume
LTE. In addition, a local treatment of excitation, such as in large
velocity gradient (LVG) or LVG-like approximations, is very useful when one
deals with expanding gas, but cannot be applied to rotating disks, in
which regions placed at large distances (comparable in fact to the disk
size) and with very different physical conditions radiatively
interact. A complex, nonlocal treatment of radiative transfer is
therefore necessary in our case.

In the treatment of the \htrececn\ line excitation, we do not include
the hyperfine structure of the rotational levels. No effects of the
partial overlap between components on radiative transfer and level
population is expected in the very optically thin rotational lines. In
the observed \jct\ line, the main components are separated by $\sim$
0.1 \kms, and the satellite components are at least ten times
weaker. The effects of the \jct\ splitting in the line profile and the
calculation of the synthetic maps are, in comparison with the
moderate S/N observations, negligible.

\section{Observational results from our ALMA maps}

Our ALMA maps of \doce\ \jtd\ and \jsc\ emission from the Red Rectangle
were presented in Bujarrabal et al.\ (2013b). In that work, we also
show maps of \trece\ \jtd; to better show the weak emission of this
line from the expanding gas far from the equator, we performed a
new image synthesis of those data, assuming natural weighting of the
visibilities (which yields a wider beam and a better sensitivity) and
degrading the spectral resolution to 0.66 \kms. As in the rest of the
images shown here, we subtracted the continuum because of the weak
line brightness (see ALMA data of the dust continuum emission in
Bujarrabal et al.\ 2013b).  The resulting \trece\ \jtd\ maps are
  shown in Fig.\ \ref{1}, in which we can see the weak emission from
  the outflow (the slight extension to high latitudes farther than
  1$''$ and following the X-shaped optical image, mostly noticeable at
  intermediate velocities, $\pm$ 2--6 \kms), almost 100 times weaker
  than the brightness peak.

In Fig.\ \ref{3}, we show our maps of \dsiete\ \jsc. As expected, the
emission is relatively weak compared with the \doce\ \jsc\ maps shown
in Bujarrabal et al.\ (2013b) and Fig.\ B.1, which reach a peak of
$\sim$ 200 K, pointing out that \dsiete\ \jsc\ is very probably
optically thin.  The outflow emission, in particular, is found to be
very faint in \dsiete\ \jsc,
at least 60 times weaker
than the peak. Meanwhile, the measured contrast in \doce\ \jsc\ is just
a factor $\sim$10.

\begin{figure*}
   \centering
   \includegraphics[width=17.3cm]{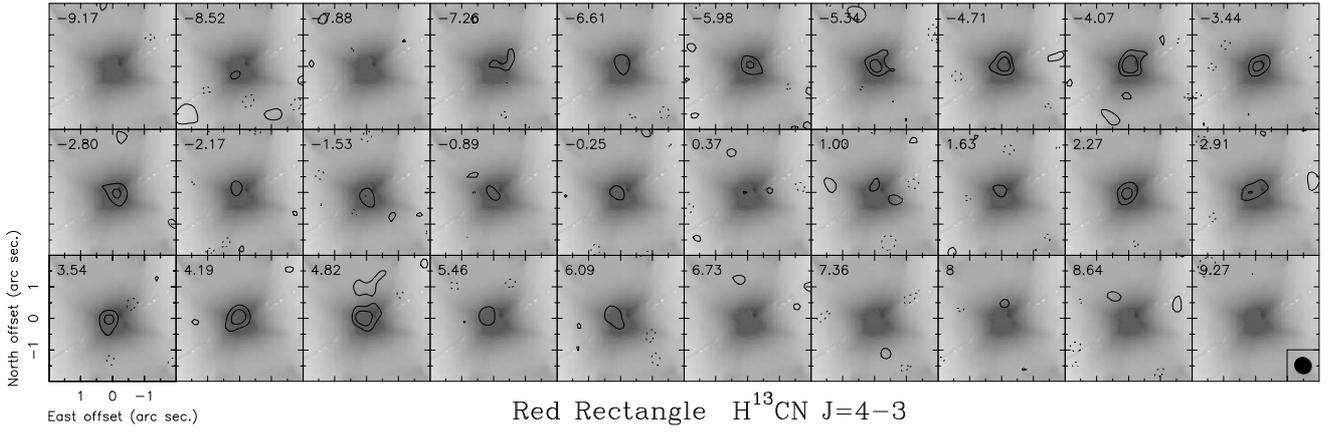}
      \caption{ALMA \htrececn\ \jct\ observations of the Red
        Rectangle. The resulting beam size (HPBW) is 0\farcss
        57$\times$0\farcss 53 (shown in the insert). The contours are
        -0.01, 0.01, 0.02, and 0.04 Jy/beam (logarithmic spacing by a
        factor 2, equivalent to 0.34, 0.68, 1.36 K). In this
        figure, the contours are separated by a factor 2.
        This is, by far, the
        weakest molecular line detected in the Red Rectangle. The
        continuum has been subtracted.} 
         \label{5}
   \end{figure*}

The \htrececn\ \jct\ maps are shown in Fig.\ \ref{5}. This is by far
the weakest molecular line detected in the Red Rectangle, this is the
only map in our observations with just a moderate S/N, of about 20. The
brightness levels, \lsim\ 1 K, are very low for a source that reaches a
brightness of $\sim$ 200 K in \doce\ \jsc\ and kinetic temperatures of
several hundred Kelvin in inner regions (see Sect.\ 5). The very broad
line with maxima at $V_{\rm lsr}$ $\sim$ $\pm$4--5 \kms\ is to be
compared with the emission of the other optically thin line,
\dsiete\ \jsc, with maxima at about $\pm$2 \kms. No trace of outflow
emission is found in our \htrececn\ \jct\ maps with a brightness
\lsim\ 7 mJy/beam, which is more than $\sim$10 times weaker than the
peak.

Further representations of our observations are shown in Figs.\ \ref{rot},
A.2, and A.3, which show in particular position-velocity (P-V) diagrams
along different cuts. We note the conspicuous Keplerian patterns in
cuts along the equator (Fig.\ A.2 and cut H00 in Figs.\ 10 and A.3) and
the tentative detection of slow rotation in gas outside the disk (H1N,
H1S, H2N, H2S in Figs.\ 10 and A.3; see Sect.\ 6). We also note the disk
velocity pattern in cuts perpendicular to the equator (inner 1$''$ in
cuts V1 and V2, Figs.\ 10 and A.3), which suggests that the rotation
velocity mostly depends on the distance to the center, not on the
distance to the axis.

   \begin{figure*}
   \centering
   \includegraphics[width=17.3cm]{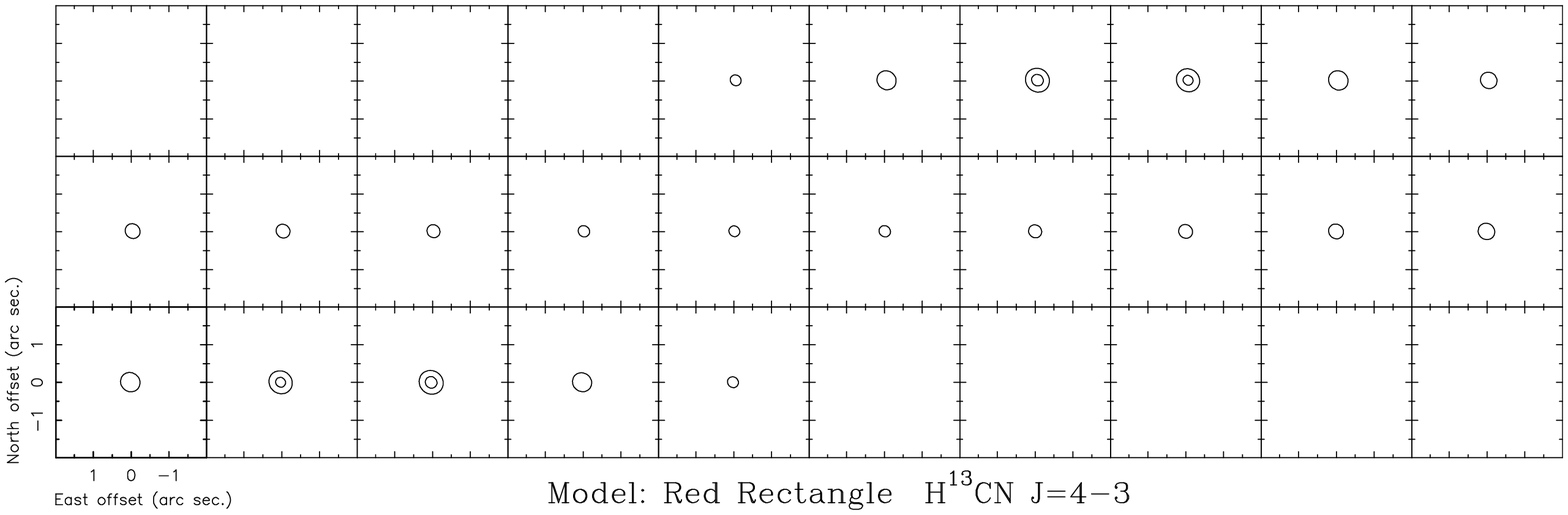}
      \caption{Theoretical maps of \htrececn\ \jct\ obtained for our
        best-fit model for comparison with ALMA maps in Fig.\ \ref{5}. The
        contours and angular and velocity units are the same as for 
        the observational data. 
   }
         \label{6}
   \end{figure*}

\section{Model fitting}

We used our numerical code (Sect.\ 3) to perform model fitting of all
maps obtained with ALMA in the Red Rectangle, namely of the lines
\doce\ \jtd\ and \jsc\ (Figs.\ 1 and 3 of Bujarrabal et al.\ 2013b and
Fig.\ \ref{map65}), \trece\ \jtd\ (Fig.\ \ref{1}),
\dsiete\ \jsc\ (Fig.\ \ref{3}), and \htrececn\ \jct\ (Fig\ \ref{5}).
We also reproduced the \doce\ and \trece\ single-dish profiles
presented in Bujarrabal \& Alcolea (2013).  As we discussed in
Sect.\ 2.3, the fraction of lost flux by over-resolution in the
interferometric observations is small, but not always negligible,
particularly in the line wings coming from the outflow. In our fitting
process, we try to fit both the total intensity obtained from
single-dish data and the maps, keeping in mind that the brightness
distributions of the mapped lines can be somewhat overestimated by our
model calculations with respect to the observations, especially for the
velocities at which the lost flux is larger.

The synthetic maps from our best-fit model are shown in
Fig.\ \ref{mod65} for the \doce\ \jsc\ (that was not fitted in our
previous paper, see observational data reproduced in
Fig.\ \ref{map65}), Fig.\ \ref{2} for the new reduction of
\trece\ \jtd\ data, Fig.\ \ref{4} for \dsiete\ \jsc, and Fig.\ \ref {6}
for \htrececn\ \jct. In Fig.\ \ref{profi} we present the new fitting of the
single-dish observations.  Other synthetic maps, to help our discussion
in the following sections on the nebula properties, are shown in App.\ B.

\subsection{Fitting of CO line maps}

\begin{figure}
   \centering
   \includegraphics[width=9cm,angle=0]{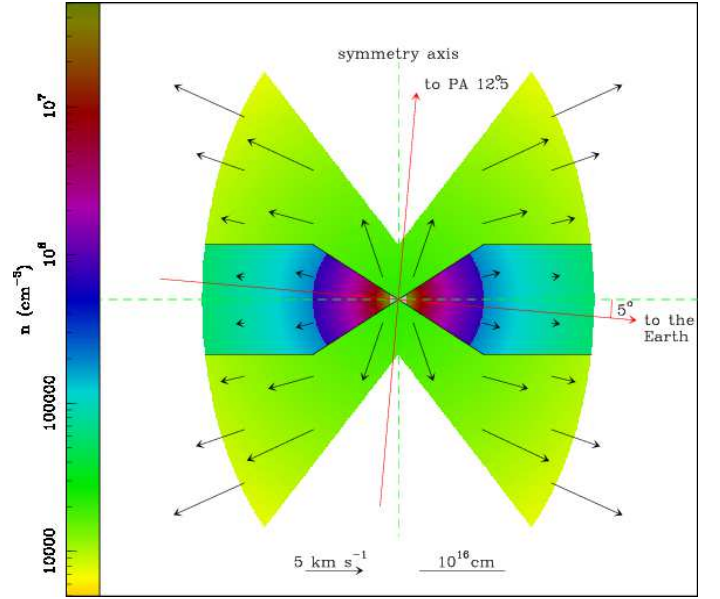}
      \caption{Structure, velocity, and density distribution in our
        best-fit model disk and outflow. We represent parameters for a
        plane perpendicular to the equator and containing the line
          of sight; therefore, the disk is seen edge-on and only
        expansion is shown. The velocity and size scales are given
          at the bottom of the figure.}
         \label{mod}
   \end{figure}

A representation of the structure, velocity field and density
distribution in our model nebula is shown in Fig.\ \ref{mod}.  In
general, the deduced nebular properties are very similar to those
presented in our previous works. The emitting regions are slightly
smaller to account for the smaller extent found in observations with
higher resolution and in optically-thin lines. In principle, the new
data are the best representation of the actual shape and the new model
should be more accurate, but we must keep in mind the flux lost in
those lines, which may lead to some underestimates of the size in the
present work. The shape of the expanding component is also somewhat
different, yielding a slightly better fitting and intuitively
accounting for the idea that the CO extent in this diffuse component
must be limited by photodissociation. The total outflow extent is
roughly compatible with photodissociation calculations by the
interstellar UV field (Mamon et al.\ 1988), which indicates that
CO-rich diffuse-enough gas expanding at about 10 \kms\ extends just a
few 10$^{16}$ cm.  In any case, the dissociation process in our complex
source is more difficult to describe than for the calculations
performed by Mamon et al.\ for a spherically symmetric AGB shell
expanding at constant radial velocity. It is in particular difficult to
be sure about the starting point and past kinetics of the outflowing
gas and the dust density in our source. Although, as mentioned, the new
convex shape tends to reproduce the observations at long distances from
the center slightly better, the exact shape of the cloud boundary is
uncertain, since it has just a moderate effect on the model
predictions.  Therefore, we do not attempt here to derive, from model
fitting or other considerations, an exact shape of the outer boundary
of the CO-rich outflow, and we just adopt that extent as
indicative of the possible real extent as it is possibly a lower limit.

The laws and parameters describing the physical conditions in the
different components of our best-fit CO-emission model are given in
Table 1 (and complemented with the graphical description in
Fig.\ \ref{mod}). As we can see in our synthetic maps (Figs.\ \ref{2},
\ref{4}, \ref{6}) and single-dish data fitting (Fig.\ \ref{profi}), the
data are correctly reproduced. From the density and size we derive, we
find a total disk mass of $\sim$ 1.3 10$^{-2}$ \ms. The mass of the
outflow is found to be $\sim$ 1.2 10$^{-3}$ \ms. These values are about
twice those found from the general analysis by Bujarrabal et al.\  (2013a),
but well within the usual uncertainty in the determination of the
parameter. The discrepancy mainly comes from the lower CO abundances
deduced here, which are $X$(\doce) $\sim$ 1.2 10$^{-4}$ and $X$(\trece)
$\sim$ 1.2 10$^{-5}$.  This value of the CO abundance is low, but it is still in
the range of what is found in post-AGB nebulae with $X$(\trece)
typically ranging between 1 and 2 10$^{-5}$ (Bujarrabal et al.\ 2001,
2015, Santander-Garc\'{\i}a et al.\ 2012, etc). The angular momentum of
our best-fit model disk is of about 2.7 10$^{52}$ g\,cm$^2$\,s$^{-1}$,
i.e.\ $\sim$ 9.1 \ms\,AU\,\kms. We recall that the analysis of lines
requiring different degrees of excitation and showing different
opacities allows us to distinguish the effects on the line intensity of
the different parameters, in particular, of the density and 
relative abundance; see some discussion on the uncertainties of the
derived values in Sect.\ 5.4.

In the comparison between the observed and predicted maps, we sometimes
find a certain overestimation of the outflow emission, particularly in
the predicted emission of the outflow in \doce\ \jsc, which shows an
X-shaped component that is wider than observed, and in
\trece\ \jtd. This could be the result of a decrease of the CO
abundance from the disk to the outflow, perhaps more important for
\trece, which is possible in view of lower density and higher exposure
to UV of the expanding component. Nevertheless, the wings of the
single-dish data, which come from the outflow, tend to be
underestimated in our fitting (Fig.\ \ref{profi}, green arrows). Any
attempt to obtain weaker outflow emission in the theoretical maps, for
instance decreasing the density or abundance in this component, affects
the angle-integrated profile, leading to a larger underestimate of the
wings of the single-dish profiles.  On the other hand, we expect a
significant loss of flux in the observed maps at relatively high
velocities ($\pm$ 4--7 \kms, Sect.\ 2.3) due to over-resolution.  We
conclude that the reason for the overestimate of the outflow emission
in our map simulations is probably the systematic loss of flux in the
observed maps. In these maps, the weak and extended emission from the
outflow, which is relatively more important at high velocities, could
be over-resolved. We think that the present fitting is a good
compromise between the fact that we know that we are losing flux in the
interferometric maps and that we must use them to derive the geometry
of the CO-rich nebula. We stress that the steep logarithmic scale we
are using tends to give a false, too pessimistic impression of the
brightness overestimate of the synthetic maps. Notably, the predicted
brightness excess for the outflow in \trece\ \jtd, shown in
Figs.\ \ref{1} and \ref{2}, affects in practice the very low first
contour (at most the second one), at a level that is about 20--50 times
weaker than the peak, and has a moderate impact on the single-dish
profiles.

The \dsiete\ \jsc\ maps are well reproduced by our model. Section 2.3
shows that the percentage of lost flux is probably much smaller in the
maps of the very compact \dsiete\ (and \htrececn) emission. A
small contribution from the outflow is seen in both the observed and
predicted maps. The model calculations confirm that the weak
\dsiete\ \jsc\ line is optically thin. Because of its low opacity
and high excitation, this line is a good probe of the very inner
regions.  We used these properties to check the strong increase in
density suggested by Men'shchikov et al.\ (2002) to explain the IR and
mm wave dust emission from this source. These authors proposed an
increase of the density by a factor \gsim\ 10$^3$ in regions closer
than 60 AU. In outer regions, the densities proposed in that paper are
similar to those found here.  However, such a density increase leads to
a significant increase of the predicted emission at velocities
expected for dense inner regions (\gsim\ 4 \kms); this is  in contrast to
our observations that show a decrease in the flux at those
velocities. Our observations thus do not support a much higher density
in the central regions. See some of our tests in App. B,
Figs.\ \ref{modb17co} and \ref{modc17co}.

\subsection{Fitting of single-dish observations of CO emission}

\begin{figure}
   \centering \rotatebox{0}{\resizebox{9cm}{!}{ 
\includegraphics{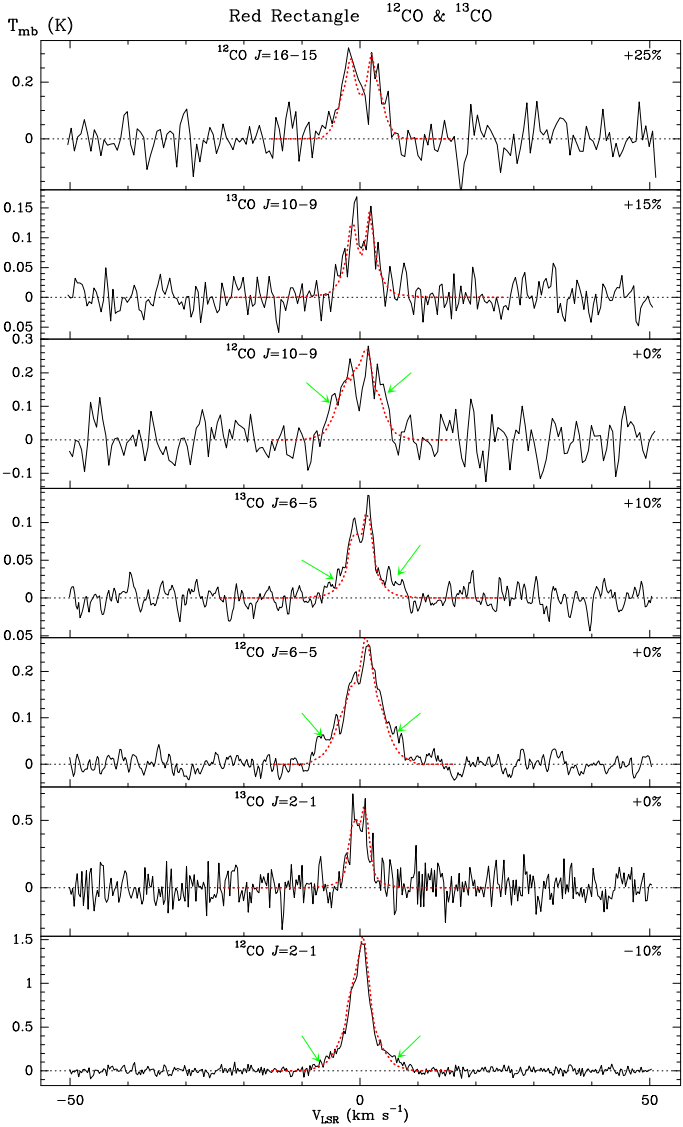}
}}
   \caption{Observed CO line profiles of the mm and FIR transitions
     (black solid line) and the predictions of our code (red dotted
     line); see Sect.\ 5.2. The line-wing emission of some lines are
     somewhat underestimated (green arrows). The values of the free
     scale parameter applied to the model predictions that account for
     uncertainties in the calibration are indicated in the upper right
     corners.}
              \label{profi}%
    \end{figure}

We also tried to reproduce single-dish mm wave and FIR observations of
the Red Rectangle; see Sect.\ 2.2. These observations were obtained
with the 30\,m IRAM radio telescope and Herschel/HIFI; see further
discussion in Bujarrabal \& Alcolea (2013). The representation and
fitting procedures for these observations are the same as in that
paper: the red points are the theoretical results and the black lines
are the observed spectra. In the comparison of the observations and the
predictions from our best-fit model (Fig.\ \ref{profi}), we included
and include now a free scale parameter (applied to the predictions)
accounting for the calibration uncertainty of the data, which is
indicated in percentage in the upper right corner of each of the
panels.

The fitting is reasonably good, requiring moderate scaling factors that are
compatible with calibration uncertainties. There is a
systematic underestimate of the line wings that come from the
(relatively fast) outflow; see green arrows in the figure. These
underestimates appear in spite of the trend to overestimate the outflow
emission in the synthetic maps; this is probably due to a certain degree
of over-resolved flux in the interferometric maps of this component
(Sect.\ 5.2); we do not try to improve in order to prevent a worse fitting
of the maps.

\subsection{Models of H\,$^{\it 13}$CN emission}

Our maps of \htrececn\ \jct\ show weak emission, with a peak of about 1
K, which is clearly indicative of very optically thin emission in
comparison with the brightness levels of other lines. The dependence of
the emission on the LSR velocity is very different from that found in
CO lines. \htrececn\ \jct\ clearly shows intense emission at velocities
of $\pm$5 \kms, which are not found in \dsiete\ \jsc; see
Figs.\ \ref{3} and \ref{5} (see also App.\ A, Fig.\ A.2).  We tried to
reproduce this behavior using the same model as for CO, and it has been
found to be impossible (App.\ B, Fig.\ \ref{modbhcn}) unless we
introduce strong changes in the HCN abundance depending on the distance
to the center. The best fitting is obtained when we assume that
\htrececn\ is abundant with a reasonable value of its relative
abundance $X$(\htrececn) $\sim$ 10$^{-9}$ only in regions closer than
60 AU (drawn in gray in Fig.\ \ref{mod}), which are precisely those in
which Men'shchikov et al.\ (2002) proposed a very strong increase of
the density; see Sect.\ 5.1. Otherwise, emission from outer regions at
lower velocities largely dominates in the maps (as actually happens for
the optically thin emission of \dsiete).

\begin{figure}
   \centering \rotatebox{0}{\resizebox{9cm}{!}{ 
\includegraphics{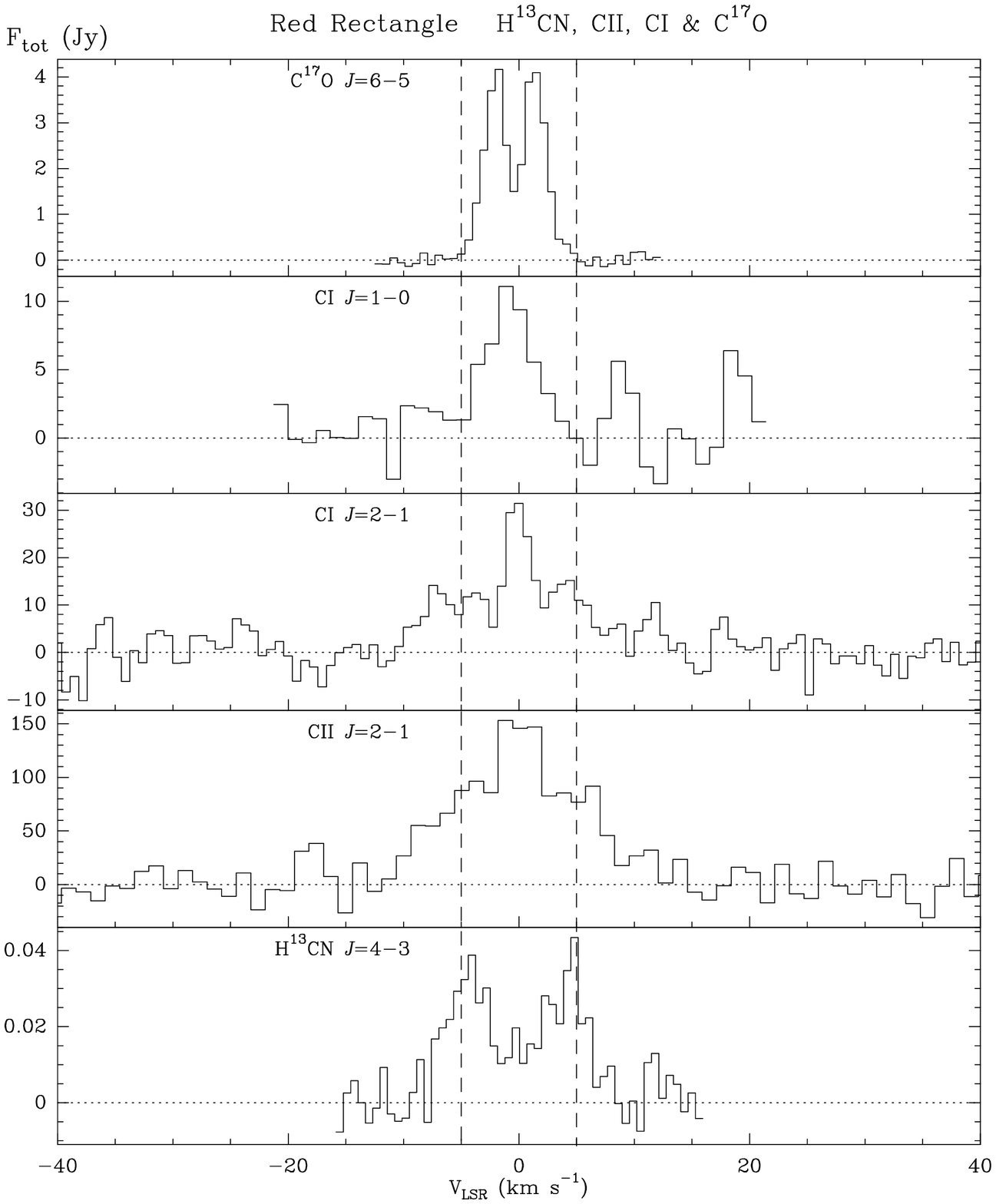}
}}
   \caption{Observed profiles of the angle-integrated emission of
     \htrececn\ \jct, CII, CI, and \dsiete\ \jct, showing the relative
     contribution of the central rapidly rotating regions. The vertical
     dashed lines approximately represent the rotation velocity in
     regions at about 60 AU from the center; see Sect.\ 5.3.}
              \label{pdr}%
    \end{figure}

The significant increase of the \htrececn\ could be due to the
development of a PDR in dense disk regions close to the central stellar
system. The stellar component of the Red Rectangle is a binary system
with a secondary star and an accretion disk emitting in the UV (Witt et
al.\ 2009; Thomas et al.\ 2013). This UV emission is known to be able
to excite a central HII region (e.g., Jura et al.\ 1997) and,
therefore, must also yield a PDR intermediate between the HII region
and the extended CO-rich disk.  Photoinduced formation of HCN has been
observed in inner regions of several disks around young stars and is
predicted by theoretical modeling of the PDR chemistry, as a result of
reactions triggered by photodissociation of stable molecules in dense
PDRs (e.g., Ag\'undez et al.\ 2008; Fuente et al.\ 2012). The presence
of photoinduced chemistry is independently supported by observations of
the C$_2$ and CH$^+$ radicals in our source (see Wehres et al.\ 2010
and Balm \& Jura 1992) and is convincingly confirmed by our single-dish
Herschel/HIFI observations of CII and CI FIR lines (Fig.\ \ref{pdr}).
CII emission is thought to originate in PDRs and is in fact the most
reliable tracer of these regions in interstellar and circumstellar
nebulae, although this emission is less clearly associated with high
densities than the production of HCN. CI is often associated with both
CO-rich gas and PDRs. As we can see in Fig.\ \ref{pdr}, the CII line
profile shows two components: a wide and intense wing that is very
similar to the \htrececn\ profiles and a central narrow feature. The
origin of the central feature is uncertain; it could come from slowly
rotating regions (perhaps the disk boundary exposed to UV photons) or
from the very inner outflow (in which axial velocity dominates). The wide
wings certainly come from the same region as the \htrececn\ emission
and are very probably tracing the PDR developed in the very inner
disk. The CI \jdu\ profile is very similar to the CII profile; in the
\juc\ line the central feature seems to dominate, but this line is very
noisy and any conclusion is uncertain. The upper panel of Fig.\ 9 shows
the difference with respect to the \dsiete\ emission, which obviously
comes from disk regions less close to the star, in spite of the high
excitation required to produce \jsc\ emission.

A very low $X$(\htrececn) in the entire disk and a sharp, very strong
increase of the density in the inner regions (as that mentioned in
Sect.\ 5.1) can explain the \htrececn\ \jct\ data, but (as also
mentioned in Sect.\ 5.1) the results would not be compatible with
\dsiete\ \jsc\ maps unless we assume that CO abundances decrease very
strongly in such inner regions. CO is often found to be depleted onto
grains in disks around young stars, but only for very low temperatures
and not in warm regions as the central disk of the Red Rectangle. On
the other hand, we proposed that \htrececn\ is abundant in inner
regions precisely because a PDR is developing in them and CO is being
photodissociated. If we assume a significant photodissociation of CO in
inner regions, models with an increase of the density at distances
shorter than about 60 AU could then yield acceptable \dsiete\ emission.
However, assuming complementary variations of the \dsiete\ abundance
and the density, by several orders of magnitude, seems to us too ad hoc
and difficult to understand.  We accordingly think that this option is
not acceptable without a significant overabundance of HCN owing to the
presence of a central PDR.
We cannot rule out the presence of significant CO photodissociation in
such inner regions in such a way that the emission of CO isotopologues
from them becomes weaker than for normal abundances. This complementary
behavior of the \htrececn\ and \dsiete\ chemistry in the inner disk
regions would be compatible with the different behavior of the observed
\dsiete\ and \htrececn\ emission.

\begin{table*}[bthp]
\caption{Structure and physical conditions in the molecular disk in the
  Red Rectangle derived from our model fitting of the CO data. The
  fitting of the \htrececn\ line requires the same structure and
  velocity field but with a peculiar abundance distribution; see
  Sect.\ 5.3. The values of the physical conditions depend on three
  geometrical parameters: the distance to the center, $r$, the
  distance to the inner conical boundary of the outflow, $p_{\rm c}$,
  and the distance to the equator, $h$. See Fig.\ \ref{mod} for a scheme of
  the density and velocity distribution. } {\tiny
\begin{center}                                          
\begin{tabular}{|l|cc|cc|cc|}
\hline\hline
 & &  &  & & & \\ 
 & \multicolumn{2}{c|}{Inner disk ($r < R_{\rm kep}$)} &  
\multicolumn{2}{c|}{Outer disk ($r > R_{\rm kep}$)} & \multicolumn{2}{c|}{Outflow} \\ 
& & & & & & \\
{Parameter}  & {Law} & { Values} &   {Law} & {Values}  &   {Law} & {Values} \\ 
& & & & & & \\
\hline\hline
  &  &  &  & & & \\
Outer radius  &   &  $R_{\rm kep}$ = 10$^{6}$
 cm &  & $R_{\rm out}$ = 2.3 10$^{16}$ cm & & \\
  &  &  &  & & & \\
\hline
  &  &  &  & & & \\
Thickness & linear & {$H(R_{\rm kep})$ = 1.3 10$^{16}$
 } cm & constant & {$H$ = 1.3 10$^{16}$} cm &ellipsoidal & \\
 & & {$H(0)$ = 6.5 10$^{15}$} cm &  & & (see Fig.\ \ref{mod}) & \\
\hline
  &  &  &  & & & \\
Tangential & $V_t \propto 1/\sqrt{r}$ & ~~$V_t(R_{\rm kep})$ = 1.5 \kms
 & $V_t \propto 1/r$ & $V_t(R_{\rm kep})$ = 1.5 \kms & & \\
velocity  & (Keplerian) & (central mass: 1.7 \ms) & (ang.\ mom.\ cons.) & & & \\
\hline
  &  &  &  & & & \\
Expansion & & 0 \kms\ & $V_{\rm exp} \propto \sqrt{a + b/r}$   
& $V_{\rm exp}(R_{\rm kep})$ = 1.6 \kms & $|V| \propto \frac{R_{\rm out} - p_{\rm c}}{R_{\rm out}} 
+ \frac{h - H/2}{3 ~10^{16}}$ & $|V_{\rm max}|$ = 12 \kms \\
velocity &  & & & $V_{\rm exp}(R_{\rm out})$ = 0 \kms & & \\
\hline
  &  &  &  &  &  & \\
 Temperature  & $T \propto 1/r^{\alpha_T}$ & $T(R_{\rm kep})$ =
   80 K & $T \propto 1/r^{\alpha_T}$ & $T(R_{\rm kep})$ = 95 K  & $T \propto 
   \left(\frac{R_{\rm kep}}{p_{\rm c} - R_{\rm kep}}\right)^{\alpha_T}$ & $T_{\rm max}$ = 500 K \\ 
 & & $\alpha_T$ = 1 & & $\alpha_T$ = 0.6  &  & $\alpha_T$ = 1 \\
\hline
&  &  &  &  &  & \\
 Gas density & $n \propto 1/r^{\alpha_n}$ & $n(R_{\rm kep})$ = 7 10$^5$ 
cm$^{-3}$ 
& $n \propto 1/r^{\alpha_n}$ & $n(R_{\rm kep})$ = 2 10$^5$ cm$^{-3}$  & 
$n \propto \left(\frac{6.4 ~10^{16} - r + R_{\rm kep}}{6.4 ~10^{16}}\right)^{\alpha_n}$ & 
$n(R_{\rm kep})$ = 1.6$\times$10$^4$ cm$^{-3}$ \\
  &  & $\alpha_n$ = 2 &  &  $\alpha_n$ = 1.5  &  &  $\alpha_n$ = 2 \\
  \hline
\end{tabular}
\begin{tabular}{|l|cc|l|}
\hline
 & & & \\
{Other parameters}  & {Law} & {Values}
 & comments \\ 
& & & \\
\hline\hline
 & & & \\
Axis inclination from the plane of the sky & & 5$^\circ$ & from optical and
CO data  \\
 & & & \\
\hline
 & & & \\
Axis inclination in the plane of the sky (PA) & & 12\fdeg 5 & from
optical and CO data \\
 & & & \\
\hline
& & & \\
Distance  & & 710 pc & various arguments (Sect.\ 1) \\
 & & & \\
\hline
 & & & \\
\doce\ relative abundance & ~~constant~~ & ~~1.2 10$^{-4}$~~  & this paper  \\
\trece\ relative abundance & ~~constant~~ & ~~1.2 10$^{-5}$~~  & this paper  \\
\dsiete\ relative abundance & ~~constant~~ & ~~6 10$^{-7}$~~  & this paper  \\
& & & \\
\hline\hline
\end{tabular}
\end{center}
}
\end{table*}

 \subsection{Uncertainties in the model fitting}

The number of parameters defining our model nebula is very large and,
in fact, not well defined, because the shape and the distributions of
the velocity, temperature, and density could admit a high variety of
laws. In addition, the amount of data to be fitted is enormous, with
maps of several lines for a high number of velocity
channels, as well as single-dish profiles. It is difficult to assign
weights to the many observational features, because of the very high
dynamic range, complex effects of flux loss, and, in general, various
degrees of uncertainty. Therefore, our fitting is just performed by
inspection of a high number of predictions and adoption of a solution
that is compatible with all observations, taking the instrumental
accuracy and various arguments on the probable nebular properties into
account.  For the same reasons, a mathematical discussion on the
uncertainties in our model fitting is very difficult.  We just estimate
the uncertainties of the main overall properties of the nebula, by
looking for variations from our best-fit values that lead to
predictions that are incompatible with the data.

The estimate of the disk dynamics is very accurate at a fraction of a
\kms, since the maps define its rotation very well. The overall
expansion velocity is also well estimated with a conservative upper
limit to the uncertainty of about $\pm$ 2 \kms, since the predicted
outflow emission velocity is clearly in contradiction with the observed
velocity extent for models with larger differences.

We have mentioned that the total extent of the molecule-rich nebula is
uncertain, in particular, because of the presence of lost flux in the
interferometric data and the lack of sensitivity to detect very weak
extended emission. It is probable that the values we give here are a
lower limit. Of course, the total size of the nebula is underestimated
because certainly it extends farther than the detected emission
(Sects.\ 1, 5.1), at least as a result of photodissociation by
interstellar UV photons. The rest of the nebula structure is well
defined by our high-resolution observations and is in fact compatible
with estimates in our previous papers. The resolution of our maps
indicate that the width of the disk, in particular, is accurate within
a 30\%.

The density and abundance uncertainties are given by the need to match
the observed intensities. In view of the significant calibration
uncertainties and lost flux, discussed before, we adopt (to
estimate uncertainties of our determinations) a conservative limit to
acceptable discrepancies of about 50\%, which could be lower for
certain velocities and lines of sight. In optically thin lines, such as
\trece\ \jdn\ and \juc\ and \dsiete\ \jsc, the changes in the
intensities are roughly proportional to the changes in the
abundance. Low-$J$, easily excited CO transitions (say $J$ $<$ 6) show
a linear dependence on the density, but high-$J$ lines roughly vary
with the square density. Therefore, the effects of both density and
abundance cannot compensate when one considers both low- and high-$J$
lines. We conclude (and quantitative calculations confirm this result) that
50\% is a reasonable limit to the uncertainty of our estimates of the
overall density and abundance distributions.

We think that the uncertainty in the temperature is lower, because
high-$J$ transitions are very sensitive to variations in the
temperature. For instance, the synthetic \dsiete\ \jsc\ brightness
decreases very strongly if we consider lower temperatures, and
\jdq\ and \jdn\ profiles also vary strongly with the temperature.  On
the other hand, opaque low-$J$ line intensities (such as \doce\ \jdu)
depend linearly on temperature, but slightly on density. Our
calculations show that variations of the characteristic temperature
larger than 
about $\pm$ 30\% with respect to the best-fitting values are in
contradiction with the observations, at least for the rotating disk;
for the outflow, uncertainties in the temperature can be higher,
$\sim$ 50\%, because of the observational issues affecting this
component.

For \htrececn, we assume the same physical conditions deduced for
CO. Those maps are relatively noisy and the structure assumed for the
very inner regions is not well established.  We discussed the extent of
the high-abundance inner region in Sect.\ 5.3: it is clear that
\htrececn\ cannot be very abundant along the whole disk, but we cannot
delimit the size and shape of the inner region rich in \htrececn\ with
high accuracy.  Only an abundance uncertainty of about a factor 2 can
be deduced in this case.

   \begin{figure*}
   \centering\rotatebox{0}{\resizebox{12cm}{!}{ 
\includegraphics{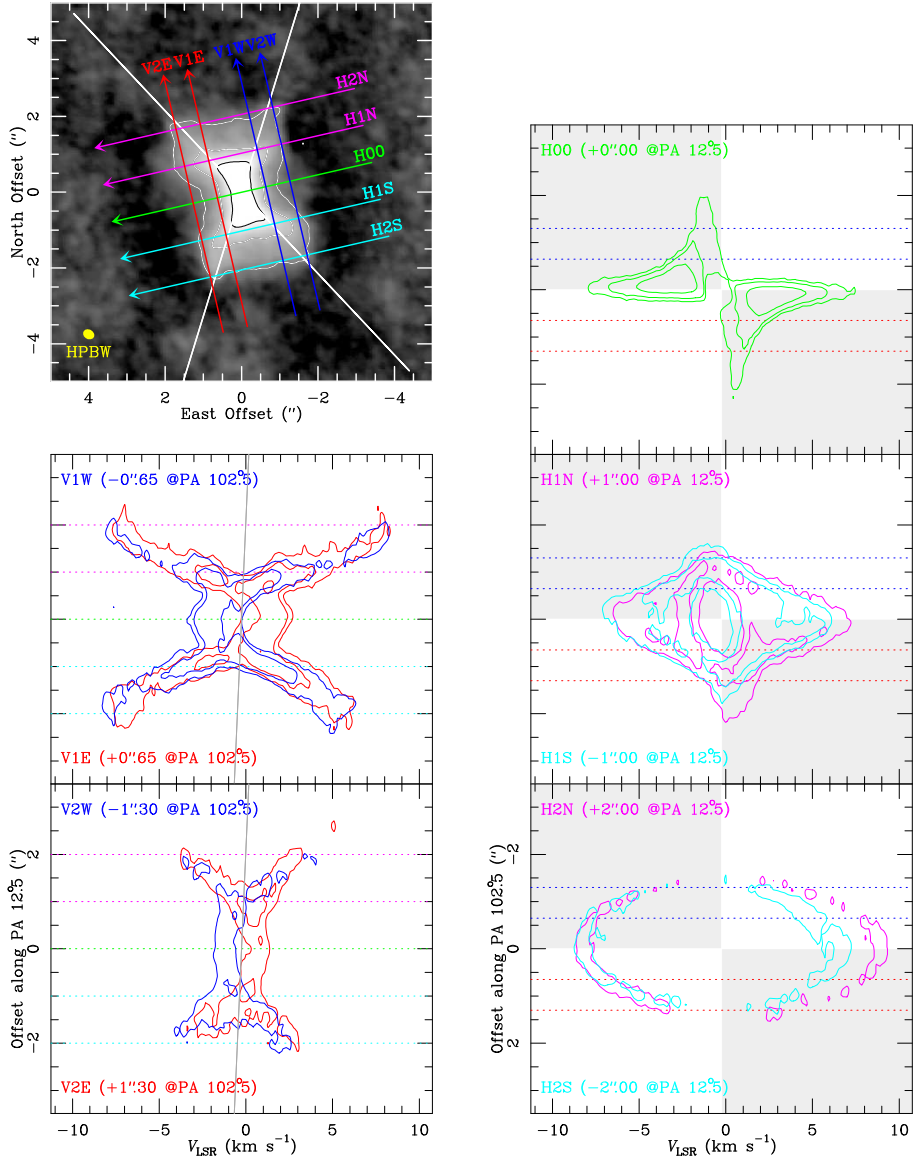}
}}
   \caption{Position-velocity diagrams of \doce\ \jsc\ for several
     cuts, indicated in the upper left panel, which represents the
     integrated flux map (the arrows indicate
       the sense of the offsets  in the cuts). The cut
     positions are also indicated, when appropriate, in
     other panels. For the integrated flux, the first contour is
     20 mJy \kms, with a logarithmic jump of a factor 3. For the P-V
     cuts, the contours are always the same as in the corresponding
     channel maps.  The dark areas in the
     horizontal cuts indicate the P-V quadrants where we expect that
     emission tends to concentrate in the case of rotation.}
              \label{rot}%
   \end{figure*}

\section{Origin of the outflowing gas: Tentative rotation of the expanding lobes}

It has been argued that the mass-loss rate of the post-AGB star in the
center of the Red Rectangle must be negligible (e.g., Van Winckel
2003), therefore, the material that forms the expanding gas detected
outside the equatorial disk should come from the disk, which acts as a
relatively stable reservoir. Moreover, the outflow structure and
velocity deduced from our model fitting strongly support this idea,
since we see gas expanding at long distances from the axis and close to
the disk, even if most of the expanding gas seems to come from the
inner, very dense disk regions.  In fact, it is known that the outer
parts of the disk shows a slow expansion, together with the obvious
rotation (to explain the asymmetry found between the negative and
positive velocities in the equatorial regions, see Bujarrabal et al.
2005, 2013b).

If the outflowing gas comes from the disk, it is expected that such a
component must share its rotation dynamics at least partially
(independent of the dominant expansion). The rotation velocity of the
lobes should be very slow because of angular momentum conservation. For
radial expansion (which is not necessarily the case here), we would
expect a decrease of the velocity proportional to the distance covered
during the expansion. For instance, if a particle leaves a region that
rotates at, say, 2 \kms, at about 5 10$^{15}$ cm, one expects a
rotation velocity of just 0.5 \kms\ at a distance of 2 10$^{16}$
cm. Such a field is obviously very difficult to detect, mostly because
expansion in those regions reaches a few \kms. We are aware of that
this reasoning is vague and uncertain because of our poor knowledge of
the path followed by the expanding gas, which can be only guessed from
the observations. In any case, some rotation should be present in the
outflowing gas.

We think that the rotation of the outflow is tentatively detected in
our maps. In Fig.\ \ref{rot}, we can see position-velocity (P-V)
diagrams along various cuts of our \doce\ \jsc\ image (Bujarrabal et
al.\ 2013b); see also Fig.\ A.3 for \doce\ \jtd. We can see that the
cuts H1N and H1S in the figure partially share the rotation pattern
seen along the equator (H00). Velocities tend to be slightly
positive/negative for positive/negative (eastward/westward) offsets,
for both north and south cuts (superposed to the expansion dynamics
dominant in these regions). The effect is small, smaller than 1 \kms,
in agreement with our discussion in the previous paragraph. As
expected, rotation is much less noticeable for cuts H2N and H2S, since
the rotation velocities must still be smaller in those cuts, but the
asymmetry can still be seen. A weak effect is also present in cuts that
are parallel to the axis, cuts V.  At distances $\sim$ $\pm$1--1.5
arcsecond from the equator, the emission is slightly bluer for VW and
redder for VE.

It is improbable that instrumental effects are responsible for these
velocity shifts. Contamination of the equatorial P-V distribution
through sidelobes is not likely because of the small ALMA beam ($\sim$
0\farcss 25 in this case). Moreover, the structure of the equatorial
P-V diagram is not reproduced at all in those along the other
horizontal cuts, in which rotation is clearly coupled with the
expansion signature (the hollow, elliptical features in the P-V
diagrams). Contamination by emission of the eventually warped outskirts
of the disk is also not likely because in that case one would expect
prominent emission at the central offset and velocity.  A similar
result is found from our maps of \doce\ \jtd, App.\ A, Fig. A.3, which
show shifts that are almost identical to those seen in \jsc. The very
comparable behavior of both \jtd\ and \jsc\ maps, despite the very
different angular resolutions (about a factor 2 better in \jsc) and
observing conditions, supports that this feature corresponds to true
rotation and not to instrumental effects.

\section{Discussion and conclusions} 

We present high-quality ALMA maps of \doce\ and \trece\ \jtd\ emission,
\doce\ and \dsiete\ \jsc\ emission, and of the \htrececn\ \jct\ line in
the Red Rectangle. The complex structure of the source, which is
basically composed of a rotating equatorial disk and a low-collimation
bipolar outflow, is confirmed (Sects.\ 4, 5). The \doce\ and
\trece\ results were already published (Bujarrabal et al.\ 2013b), but
we present here the results of a new reduction of the
\trece\ \jtd\ maps that show emission from the outflow, not obviously
seen in our previous publication. We also present Herschel/HIFI
observations of CI and CII line emission from this source, which are
compared with the CO and \htrececn\ lines. A detailed model fitting of
all the molecular line data, including previously published single-dish
observations of several rotational lines of CO, was performed using a
sophisticated code that includes accurate nonlocal treatment of
radiative transfer (see Sects.\ 3 and 5); the main properties of the
model nebula are summarized in Table 1 and Fig.\ \ref{mod}. The
existence of high-quality data on low- and high-excitation lines, also
showing different degrees of opacity, allows us to study the nebular
components in detail. This is the case thanks, in particular, to a more
reliable separation of the effects of density and abundance on the
calculated intensities (Sect.\ 5.4).

\subsection{Total mass of the disk and outflow}

From the density and size we derive from our modeling, we find a total
mass of the disk of $\sim$ 1.3 10$^{-2}$ \ms, while the mass of the
outflow is $\sim$ 1.2 10$^{-3}$ \ms. These values are about two times
higher than those found from the general and simpler analysis by
Bujarrabal et al.\ (2013a), the discrepancy mainly comes from the lower
CO abundance deduced here, i.e., $X$(\doce) $\sim$ 1.2 10$^{-4}$ and
$X$(\trece) $\sim$ 1.2 10$^{-5}$. These values of the abundance (as well
as those used in our previous paper, 2 10$^{-4}$ and 2 10$^{-5}$,
respectively) are compatible with usual estimates of this parameter.
The relatively high values of the density and temperature we adopt here
are necessary to explain the high excitation found in the disk and
outflow, which show intense emission in high-$J$ lines. Therefore, the
abundance must at the same time decrease to yield the observed
intensities.

The angular momentum of our best-fit model disk is of about 2.7
10$^{52}$ g\,cm$^2$\,s$^{-1}$, i.e.\ $\sim$ 9.1 \ms\,AU\,\kms. This value
is comparable to (or somewhat smaller than) the angular momentum of a
subsolar companion with the orbital period of the Red Rectangle system,
which is 320 days. Therefore, if we assume that the disk angular
momentum comes from that of the binary system, we conclude that a good
deal of the original angular momentum of the system was (in some way)
transferred to the disk and that the distance between the stars
decreased by a factor of a few to tangentially accelerate the disk
material. Nevertheless, the disk of the Red Rectangle seems to be
particularly massive and extended, compared with other similar objects
for which data exist (Sect.\ 1), therefore, it is possible that the
formation of the Keplerian disks in these other objects have been in
general less demanding.

\subsection{Structure and chemistry of the inner Keplerian disk}

Thanks to our high-sensitivity observations of the optically thin,
high-excitation lines \dsiete\ \jsc\ and \htrececn\ \jct\ we are able
to probe the very inner regions of the Keplerian disk. We recall that
the well-defined Keplerian dynamics of these regions allows the
identification of emission from central regions, rotating at high
velocity, even if their extent is at the limit of the telescope
resolution.

The \dsiete\ line is useful to study regions even closer than 60 AU
from the central star, which rotate at velocities \gsim\ 5 \kms. The
good fit of the observations with our model shows that the density and
temperature of these inner regions still follow the smooth function we
find for outer regions (Sect.\ 5.1). In particular, we do not find any
sign of the important increase of the density proposed for such very
inner regions from studies of the IR dust emission of the Red
Rectangle. Such an increase should have led to very significant
emission at velocities farther than $\pm$ 5 \kms, which is not
observed. Although CO is a very stable molecule and chemical effects
hardly change its abundance, we discuss below (see also Sect.\ 5.3)
that we cannot exclude that photodissociation is significant in the
very inner disk regions.

The \htrececn\ \jct\ line does show an important emission at such high
velocities. The only way to reproduce the data is to assume that
\htrececn\ is significantly abundant only in the central regions of the
disk, precisely closer than 60 AU. We think, following results that are
well established in the interstellar medium, that HCN is efficiently
formed only in a central PDR (Sect.\ 5.3). The stellar component of the
Red Rectangle is known to emit enough UV to excite a central PDR.  See
further arguments supporting the presence of photoinduced chemistry in
Sect.\ 5.3, including the detection of CII and CI FIR lines, which are
the best tracers of PDRs, that are observed in the Red Rectangle and
show a wide spectral feature that corresponds to emission from this
inner disk (plus a central narrow feature of uncertain origin).

According to this scenario, we further speculate that PAHs and other
C-bearing macromolecules would be efficiently formed and excited
because of the strong UV radiation in our source.  PAHs are known to be
particularly abundant in PDRs and its emission is very intense in them,
they are in fact a very good tracer of interstellar PDRs (see Pilleri
et al.\ 2015, Vicente et al.\ 2013, Cox et al.\ 2016, and references
therein). The formation of PAHs can result from the destruction of
grains by UV photons, but they can be also formed by bottom-up
chemistry in dense gas if atomic carbon is very abundant due to CO
photodissociation, as has been discussed precisely for the case of
O-rich post-AGB nebulae by, e.g., Cox et al.\ (2016) and Guzman-Ramirez
et al.\ (2011, 2014).  The X-shaped structure seen in the optical image
of the Red Rectangle is known to be emission of PAHs and other
(probably carbonaceous) macromolecules; see Sect.\ 1 and Cohen et
al.\ (2004). We suggest that PAHs may be efficiently formed in the
inner disk regions and then expelled with the gas outflow.  This
result, if confirmed, would be practically independent of the possible
change in the dominant chemistry of the star, from O-rich to C-rich
(Sect.\ 1), since PAH emission can be intense even in O-rich PDRs.  In
Fig.\ \ref{mod}, we also show the suggested central {\it PAH factory}
(the small butterfly-like structure, in gray, with a radius of about 60
AU), in which all PAHs seen in the optical would be formed and in which
HCN is much more abundant than in the rest of the disk.

\subsection{Mass loss from the disk and its lifetime}

We have argued (Sects.\ 1, 6) that the outflowing component detected in
the Red Rectangle is probably formed of gas extracted from the disk, in
view of the structure and velocity field found in this low-velocity
outflow, the possible detection of rotation in it, and some general
properties of our sources.  The ejection of disk material could be due
to interaction with the post-AGB axial jet or  low-velocity
magnetohydrodynamical ejection of material from the extended
disk.

In addition, we derive a decrease in the density between the disk and
the outflow by a factor $\sim$ 10. With the value of the outflow
density close to the boundary of both components and the velocities
derived for the expanding gas, and taking the inclination of the
velocity with respect to the plane of the disk into account, we can
estimate a typical mass-loss rate per surface unit. The value of this
parameter at a typical distance to the axis equal to $R_{\rm kep}$, for
instance, is \mloss\ $\sim$ 1.2 10$^{-6}$
gr\,cm$^{-2}$\,yr$^{-1}$. This rate is of course uncertain, since the
velocity field very close to the disk is very difficult to derive from
the data. The surface density of the disk at that point is $M$ $\sim$
10$^{-2}$ gr\,cm$^{-2}$. Therefore, the (future) lifetime of the disk
derived at this typical distance is $t_{\rm disk}$ $\sim$ 8000 yr. A
similar value is found if we compare the total masses of the disk and
the outflow taking into account the formation time of the observed
outflow of $\sim$ 1000--1500 yr. The outflow lifetime can be derived
both from direct inspection of the velocities in our model and from the
centroid of the detected expansion in our P-V cuts in the directions
V2E, V1E, V1W, and V2W (see the almost vertical gray lines in
Figs.\ \ref {rot} and A.3). The presence of outer outflow components
not detected in CO emission, Sect.\ 1, does not significantly change
this conclusion, since a higher total mass would be roughly compensated
by a longer outflow lifetime.

The relative mass-loss rates are slightly lower/higher for
shorter/longer distances to the center; in the inner regions where
rotation is purely Keplerian, which contain about 75\% of the disk mass
in our disk model, lifetimes estimated in this way are slightly longer,
and more uncertain, since the outflow properties of these very inner
outflow regions are poorly probed by our data.  We must also take into
account that the outer parts of the disk are in slow expansion, which
should lead to mass loss of the purely Keplerian disk. We estimate a
total mass loss of the purely Keplerian regions due to this process of 
$\sim$ 10$^{-6}$ \my, which, compared to an inner disk mass of $\sim$
0.01 \ms, would lead to a lifetime of about 10000 yr. Therefore, the
lifetime of the central regions of the disk is probably comparable or
slightly higher than the representative value given above.

We therefore deduce a typical lifetime of the Red Rectangle disk
$t_{\rm disk}$ $\sim$ 8000 yr, probably $\sim$ 10000 yr for the inner
disk regions. We can obtain similar results for other similar
NIR-excess post-AGB stars; we stress that they are usually less well
studied than the Red Rectangle and therefore the estimates for them are
uncertain. In 89 Her (Bujarrabal et al.\ 2007, 2013a), the mass of both
the disk and the outflow are similar at present. The hourglass-like
outflow was formed about 3500 yr ago. If the mass-loss rate from the
disk does not significantly vary during the post-AGB phases, we can
expect a total disk lifetime of about 7000 yr, which is smaller than
the disk lifetime we found in the Red Rectangle. However, other
objects, like AC Her, show a very faint outflow. The outflow was not
detected in AC Her in CO \jdu\ at a level of about 1/15 of the peak
brightness; see Bujarrabal et al.\ (2015). In the maps of the Red
Rectangle obtained by Bujarrabal et al.\ (2005) for that same
transition and with a roughly similar relative resolution, we can see a
hint of outflow emission at a level of about 1/8 of the
peak. Therefore, the contrast between the densities in the disk and
outflow must be \gsim\ 2 times lower in AC Her. Assuming the other
properties for both nebulae (in particular the outflow velocity and
disk width) to be similar, we can conclude that the lifetime of the
disk around AC Her is \gsim\ 2 times higher than that of the Red
Rectangle.

The lifetime of the preplanetary nebula phase is of about 1000 yr, as
deduced from data of the massive usually observed nebulae. The total
lifetime of the planetary nebula phase is usually assumed to be of
about 20000 yr, after which the shells would be difficult to
detect. However, the Red Rectangle and similar objects show a
relatively low mass and their central stars are still cool
(Bujarrabal et al.\ 2013a and references therein), so their evolution
could be relatively slow. In fact, the Red Rectangle has already spent
at least 2000 yr in a post-AGB stage similar to its present one, in
view of the extent and velocity of its bipolar nebula. As mentioned
before, the hourglass-like nebulae around 89 Her was formed about 3500
yr ago. Therefore, we can expect that the preplanetary nebula phase
in these objects lasts significantly more than the above value of 1000
yr, characteristic of high-mass preplanetary nebulae.

Given the range of disk lifetimes we derive for our low-mass post-AGB
objects, we conclude that the disk mass loss in them, which is supposed
to be responsible for the outflowing gas, is the main limit to the
Keplerian disk lifetime (at least, to the gas-rich disk
lifetime). Therefore, the disk will survive long enough to be
detectable during most or all of their post-AGB (preplanetary) phases.
We think that at least some of them will survive during the planetary
nebula phase provided that the objects we are dealing with here will
evolve to form planetary nebulae and white dwarfs following, more or
less, our standard ideas on the post-AGB evolution.

\begin{acknowledgements}
This work has been supported by the Spanish MICINN, program CONSOLIDER
INGENIO 2010, grant ``ASTROMOL" (CSD2009-00038), by the Spanish MINECO,
grants AYA2012-32032 and FIS2012-32096, and by the European Research
Council (ERC Grant 610256: NANOCOSMOS). We used the Cologne Database
for Molecular Spectroscopy, for our discussion on the
\htrececn\ hyperfine structure, and the LAMBDA database, to obtain the
collisional transition rates.
\end{acknowledgements}


\newpage

\appendix

\section{Further observational results and details on data analysis}

   \begin{figure}{ht}
   \centering \rotatebox{0}{\resizebox{7cm}{!}{ 
\includegraphics{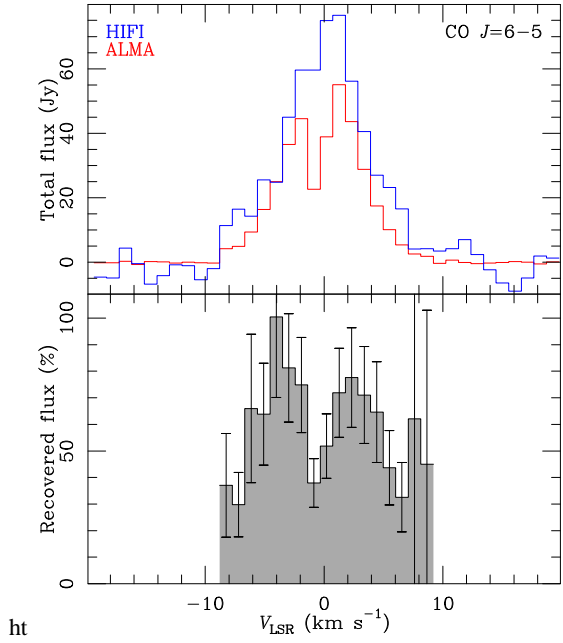}
}}
   \caption{Comparison of the \doce\ \jsc\ single-dish profile and the
     angle-integrated flux detected in our ALMA maps.}
              \label{lost}%
   \end{figure}

In the upper panel of Fig.\ A.1,  we show the flux profile observed with
Herschel/HIFI (black histogram) from Bujarrabal \& Alcolea (2013) and
the total (angle-integrated) flux detected in our ALMA maps (Bujarrabal et
al.\ 2013b). In the lower panel, we can see the fraction of the flux
actually recovered in the interferometric process, which is found to be
high, except for the central channel and intermediate velocities 
$\pm$ 5--8 \kms.

Fig.\ A.2 shows cuts along the equator for all the lines detected and
mapped with ALMA in the Red Rectangle.  We note the difference
between the \htrececn\ emission, which is proposed to come only from
very inner regions in fast rotation, and those of the CO lines,
including the optically thin \dsiete\ \jsc.

Fig.\ A.3 shows P-V diagrams for the different cuts shown in the upper
left panel for the \doce\ \jtd\ line. The asymmetry in the cuts
(particularly in the H1N and H1S ones) tentatively show the presence of
rotation in the outflowing gas. The results are very similar to that
found in the \jsc\ transition, Fig.\ \ref{rot}, showing that
instrumental effects can hardly be responsible for the observed
asymmetry. The dark areas in the horizontal cuts indicate the P-V
quadrants where we expect that emission tends to concentrate in the
case of rotation.

   \begin{figure}
   \centering \rotatebox{0}{\resizebox{7.5cm}{!}{ 
\includegraphics{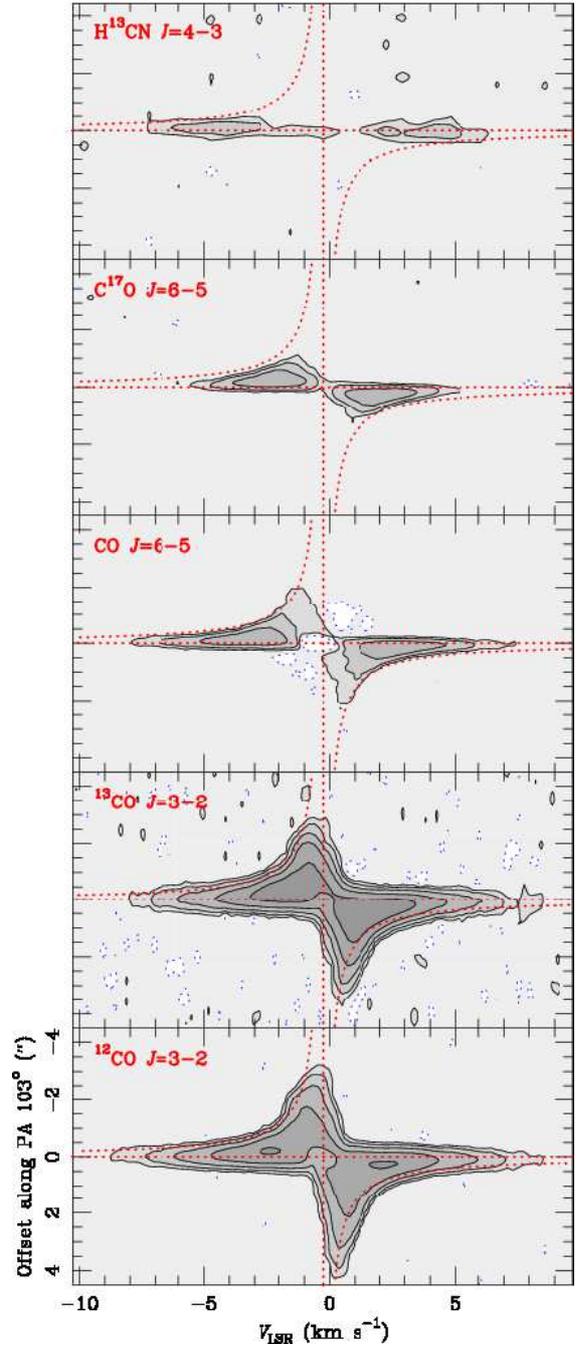}
}}
   \caption{Position-velocity diagrams found along the equatorial
     direction of the lines observed with ALMA in the Red
     Rectangle. Contours are, as usual, the same as in the corresponding
     channel maps. To help in the comparison of the different
       diagrams, we show (red dotted lines) approximate indications of
       the central position and systemic velocity of the nebula
       (horizontal and vertical lines) 
       and of        the  
       distributions of the emission of gas in Keplerian rotation (hyperbolas).}
              \label{}%
    \end{figure}
   

   \begin{figure*}
   \centering \rotatebox{0}{\resizebox{12cm}{!}{ 
\includegraphics{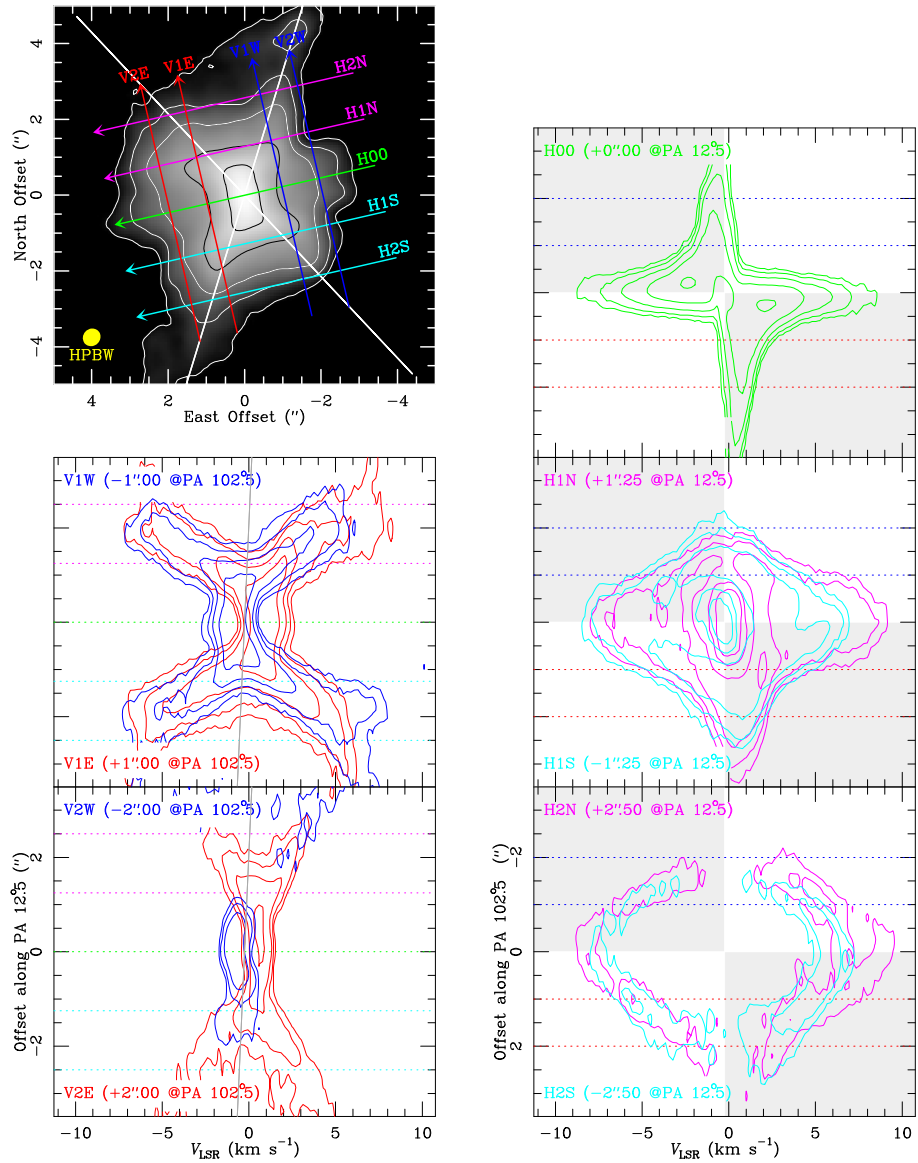}
}}
   \caption{Position-velocity (P-V) diagrams of \doce\ \jtd\ for
     several cuts, indicated in the upper left panel, which represents
     the integrated flux map; the arrows indicate the sense of the
     offsets in the cuts. The cut positions are also indicated, when
     appropriate, in other panels. For the integrated flux, the first
     contour is 60 mJy \kms\ with a logarithmic jump of a factor 3. For
     the P-V cuts, the contours are always the same as in the
     corresponding channel maps. The dark areas in the horizontal cuts
     indicate the P-V quadrants where we expect that emission tends to
     concentrate in the case of rotation.}
              \label{spectra}%
    \end{figure*}

  \section{Further details on model calculations: changes in the
     properties of the very inner disk}

   \begin{figure*}
   \centering
   \includegraphics[width=17.3cm]{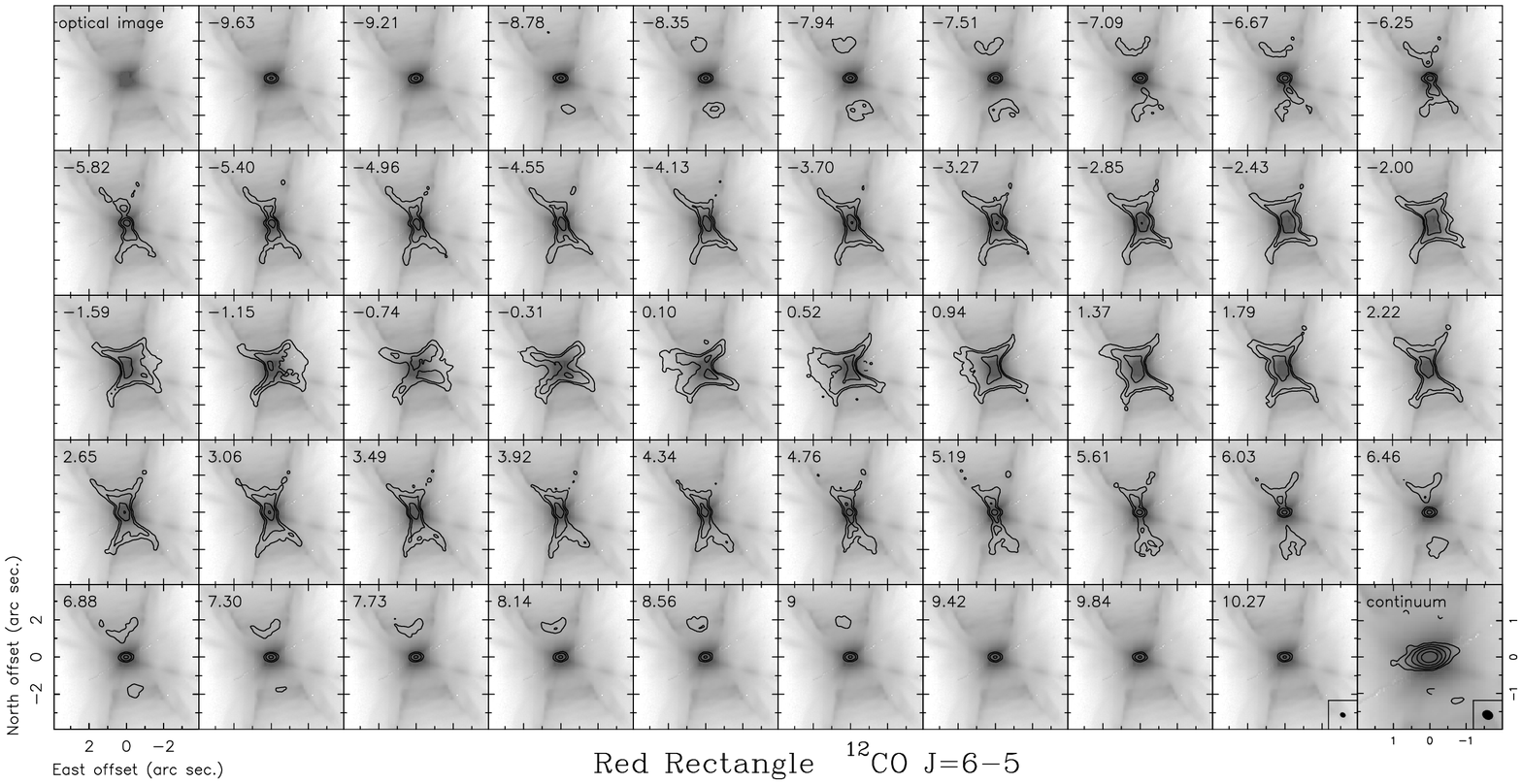}
      \caption{ALMA maps of  \doce\ \jsc, from  Bujarrabal et
        al. (2013b). This figure is reproduced here just to allow
        comparison with theoretical predictions,
        Fig.\ \ref{mod65}. Continuum is not subtracted in this case
        because of the intense line emission.}
         \label{map65}
   \end{figure*}

   \begin{figure*}
   \centering
   \includegraphics[width=17.3cm]{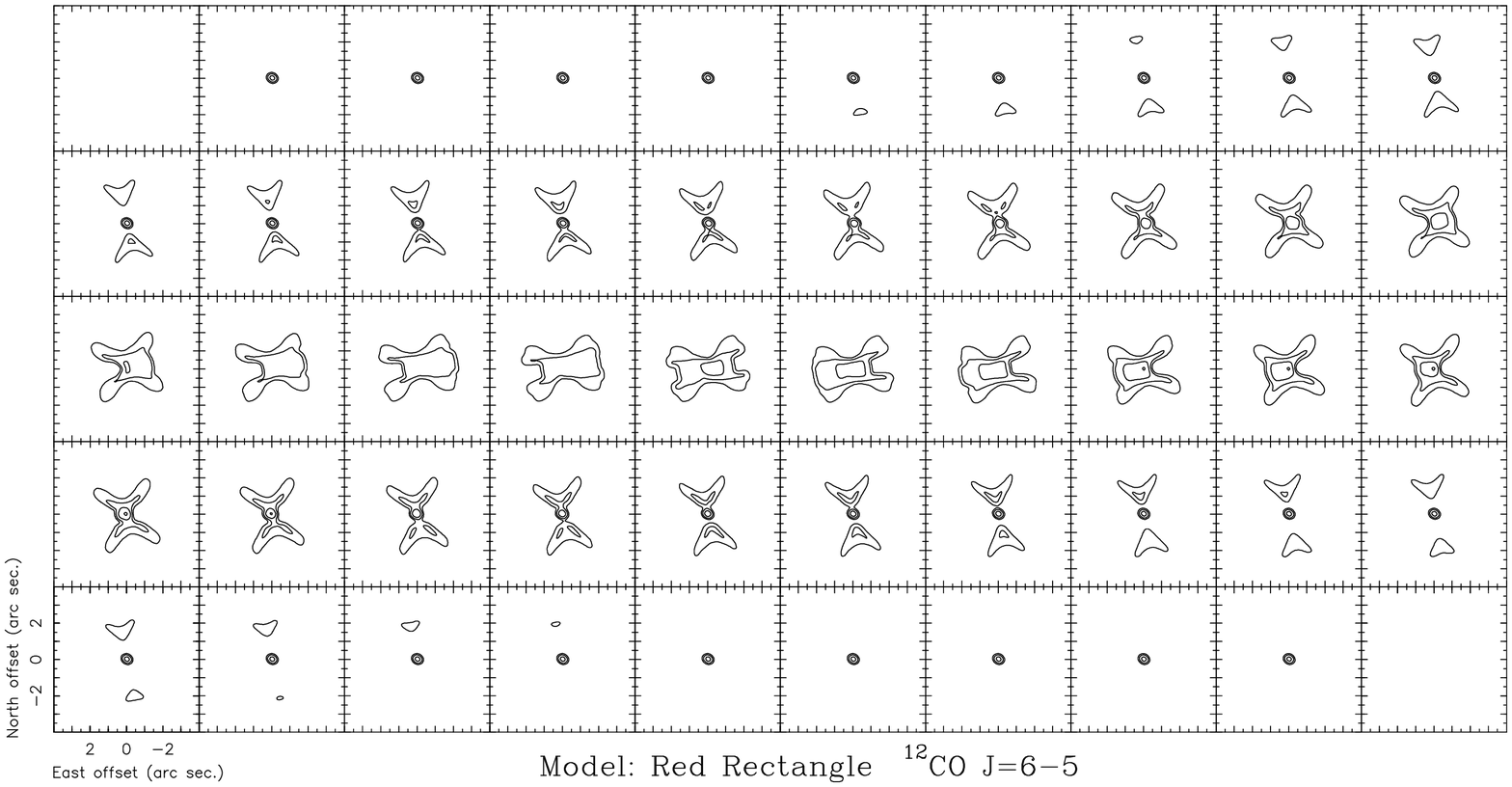}
      \caption{Theoretical maps of \doce\ \jsc\ obtained for our
        best-fit model, to be compared with ALMA maps published by
        Bujarrabal et al. (2013b), Fig.\ 3 in that paper and
        Fig.\ \ref{map65} above, in which continuum was not
        subtracted. The contours and angular and velocity units are the
        same as in the figure that shows the observational data.}
         \label{mod65}
   \end{figure*}

   In Figs.\ \ref{map65} and \ref{mod65}, we show the
   observed maps of  \doce\ \jsc\ (reproduced from Bujarrabal et
   al.\ 2013b) and the predictions from our model for this line, respectively. 

\begin{figure*}
\centering 
   \includegraphics[width=17.3cm]{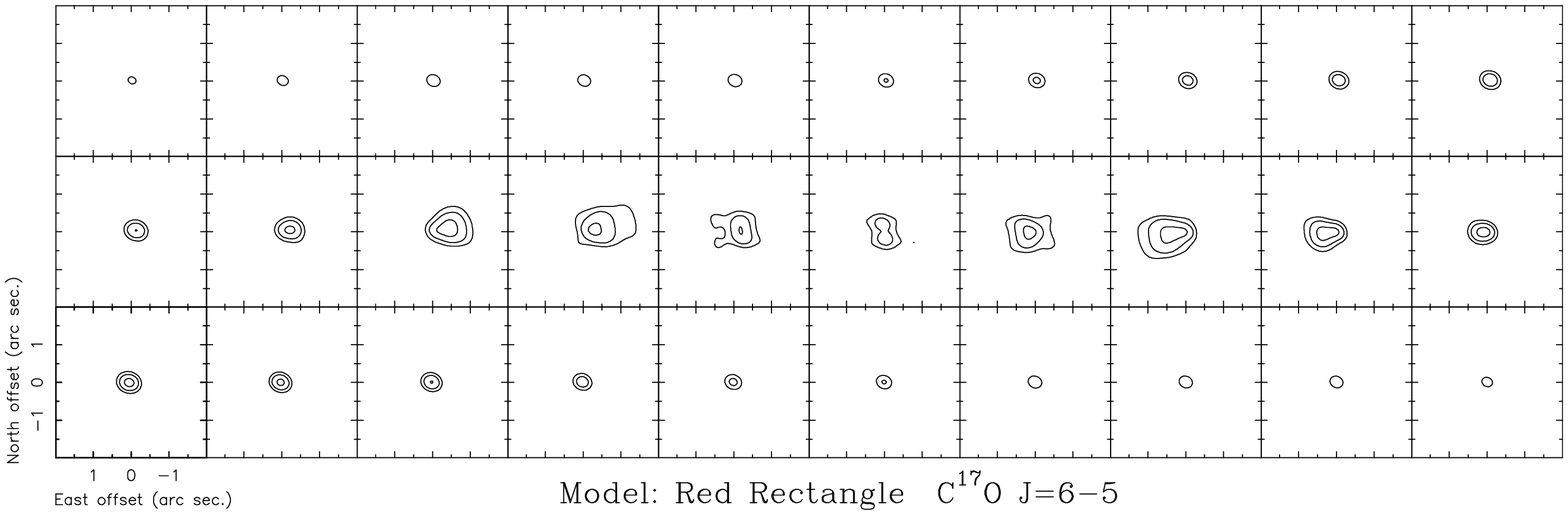} 
   \caption{Predictions of \dsiete\ \jsc\ maps assuming a
     conservative increase
     in the density of the inner disk regions by a factor 100. To be
     compared with Figs.\ \ref{3} and \ref{4}.}
              \label{modb17co}%
    \end{figure*}

\begin{figure*}
\centering 
   \includegraphics[width=17.3cm]{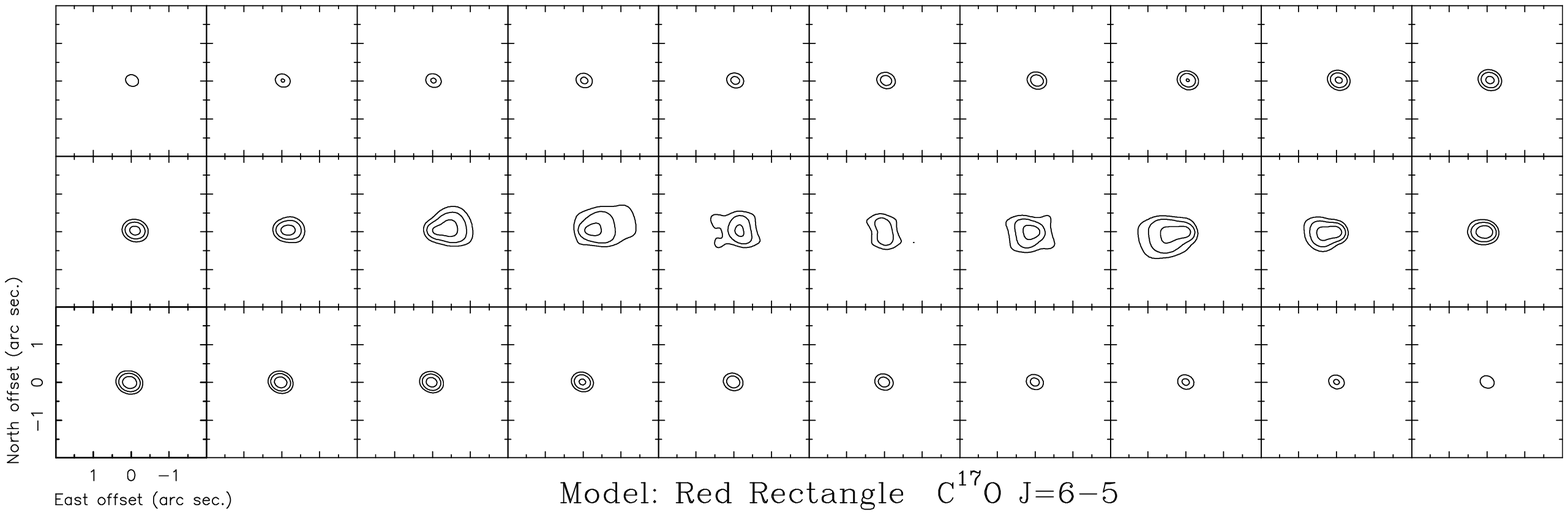} 
   \caption{Predictions of \dsiete\ \jsc\ maps assuming an increase
     in the density of the inner disk regions by a factor 1000. To be
     compared with Figs.\ \ref{3} and \ref{4}.}
              \label{modc17co}%
    \end{figure*}

   Figs.\ \ref{modb17co} and \ref{modc17co} show the predicted maps
   per velocity channel of \dsiete\ \jsc\ for increases by,
   respectively, factors of 100 and 1000 of the density in disk regions
   closer than 60 AU. A comparison of these figures with our Figs.\ 3
   and 4 shows that such a very dense central region would lead to
   emission that is  too
   intense at moderate velocities over $\pm$ 5 \kms.

\begin{figure*}
   \centering 
   \includegraphics[width=17.3cm]{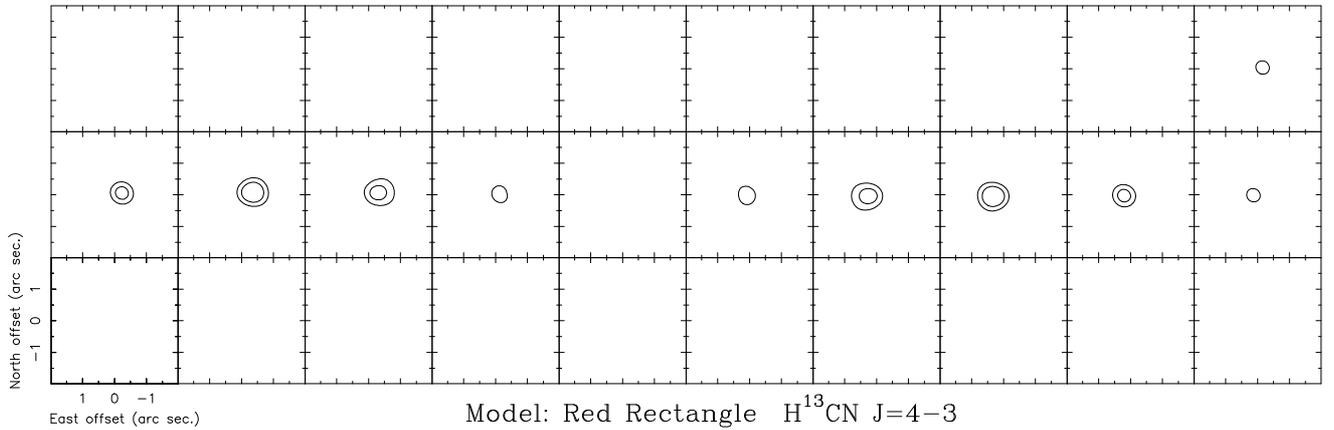}
   \caption{Predictions of \htrececn\ \jct\ maps assuming a constant
     abundance across the disk; we took in this case $X$(\htrececn) =
     10$^{-10}$, which is smaller than in our final model to avoid
     values of the total intensity that are  too
     strong. To be compared with Figs.\ \ref{5}
     and \ref{6}.}
              \label{modbhcn}%
    \end{figure*}

  Fig.\ \ref{modbhcn} shows the expected disk emission in
  \htrececn\ \jct\ emission for a constant \htrececn\ relative
  abundance of 10$^{-10}$, for comparison with Figs.\ \ref{5} and
  \ref{6}. We can see that predictions cannot match the high
  \htrececn\ emission at moderate velocities. We have chosen a smaller
  abundance than in our best-fit model by a factor 10 because the
  larger emitting region in the case presented in Fig.\ \ref{modbhcn}
  would yield to an emission that is too intense at all velocities and
  difficult to compare with the observations.

\end{document}